\documentclass[pre,twocolumn,groupaddress,showpacs,showkeys,preprintnumbers,floatfix]{revtex4-1}

\usepackage{etex}
\usepackage[mathscr]{eucal}
\usepackage{ifpdf}
\usepackage[pagebackref]{hyperref}
\usepackage{dcolumn}
\usepackage{url}
\usepackage{amsmath}
\usepackage{amscd}
\usepackage{amsfonts}
\usepackage{amssymb}
\usepackage{amsthm}
\usepackage{bigints}
\usepackage{xfrac}
\usepackage{braket}
\usepackage{bm}   
\usepackage{bbm}
\usepackage[abs]{overpic}
\usepackage{verbatim}
\usepackage{stmaryrd}
\usepackage{xcolor}
\usepackage{setspace}
\usepackage{multirow}
\usepackage{tikz}
\usetikzlibrary{arrows}
\usetikzlibrary{automata}
\usetikzlibrary{positioning}
\usetikzlibrary{plotmarks}
\usepackage{pgfplots}
\pgfplotsset{compat=newest}

\usepackage{graphicx}
\usepackage[caption=false]{subfig}

\usepackage{hyphenat}
\tikzstyle{vaucanson}=[
  node distance=2.5cm,
  bend angle=15,
  auto,
  every loop/.style={},
  every edge/.style={->,draw=black,line width=1.2,>=latex,shorten <=1pt,shorten >=1pt},
  every state/.style={draw=black,line width=2,font=\large},
  loop right/.style={right,out=22,in=-22,loop},
  loop above/.style={above,out=112,in=68,loop},
  loop left/.style={left,out=202,in=158,loop},
  loop below/.style={below,out=292,in=248,loop},
  loop above right/.style={right,out=67,in=23,loop},
  loop above left/.style={left,out=157,in=113,loop},
  loop below left/.style={left,out=247,in=203,loop},
  loop below right/.style={right,out=337,in=293,loop},
]

\theoremstyle{plain}    
\theoremstyle{plain}    
\theoremstyle{plain}    
\theoremstyle{plain}    
\theoremstyle{plain}    
\theoremstyle{plain}    
\theoremstyle{plain}    
\theoremstyle{plain}    
\theoremstyle{plain}    
\theoremstyle{plain}    
\theoremstyle{plain}    \newtheorem{Def}{Definition}
\theoremstyle{plain}    
\theoremstyle{plain}

\newcommand{\eM}     {\mbox{$\epsilon$-machine}}
\newcommand{\eMs}    {\mbox{$\epsilon$-machines}}
\newcommand{\EM}     {\mbox{$\epsilon$-Machine}}

\newcommand{\MeasAlphabet}  {\mathcal{A}}
\newcommand{\MeasSymbol}   { {X} }
\newcommand{\meassymbol}   { {x} }

\newcommand{\Past} { \smash{\overleftarrow {\MeasSymbol}} }
\newcommand{\past} { \smash{\overleftarrow {\meassymbol}} }
\newcommand{\pastprime} { {\past}^{\prime}}
\newcommand{\Future}   { \smash{\overrightarrow{\MeasSymbol}} }

\newcommand{\CausalState}   { \mathcal{S} }

\newcommand{\causalstate}   { \sigma }
\newcommand{\CausalStateSet}    { \boldsymbol{\CausalState} }
\newcommand{\AlternateState}    { \mathcal{R} }

\newcommand{\alternatestate}    { \rho }

\newcommand{\AlternateStateSet} { \boldsymbol{\AlternateState} }

\newcommand{\Prob}      {\Pr} 

\newcommand{\Cmu}       {C_\mu}
\newcommand{\hmu}       {h_\mu}
\newcommand{\EE}        {{\bf E}}
\newcommand{\TI}        {{\bf T}}
\newcommand{\SI}        {{\bf S}}

\newcommand{\PC}        {\chi}

\newcommand{\ProcessAlphabet}   {\MeasAlphabet}

\newcommand{\forward}{+}
\newcommand{\reverse}{-}
\newcommand{\forwardreverse}{\pm} 

\newcommand{\FutureCausalState} { {\CausalState}^{\forward} }

\newcommand{\PastCausalState}   { {\CausalState}^{\reverse} }

\newcommand{\one}{\mathbf{1}}

\newcommand{\lastindex}[2]{
  \edef\tempa{0}
  \edef\tempb{#2}
  \ifx\tempa\tempb
    \edef\tempc{#1}
  \else
    \edef\tempa{0}
    \edef\tempb{#1}
    \ifx\tempa\tempb
      \edef\tempc{#2}
    \else
      \edef\tempc{#1+#2}
    \fi
  \fi
  \tempc
}

\newcommand{\MOrder}{R}

\newcommand{\bmu}{b_\mu}

\newcommand{\I}{\mathbf{I}}

\newcommand{\CSjoint}[1][,]{
   \edef\tempa{:}
   \edef\tempb{#1}
   \ifx\tempa\tempb
      \ensuremath{\FutureCausalState\!#1\PastCausalState}
   \else
      \ensuremath{\FutureCausalState#1\PastCausalState}
   \fi
}

\newif\ifpm
\edef\tempa{\forwardreverse}
\edef\tempb{\pm}
\ifx\tempa\tempb
   \pmfalse
\else
   \pmtrue
\fi

\renewcommand{\H}{\operatorname{H}}
\renewcommand{\I}{\operatorname{I}}
\newcommand{\z}{z}  
\newcommand{\R}{R}  
\newcommand{\EEsp}{\EE(\omega)}
\newcommand{\SIsp}{\SI(\omega)}
\newcommand{\Psp}{P(\omega)}

\colorlet {R_color}    {blue}
\colorlet {k_color}    {black!30!green}

\def\clap#1{\hbox to 0pt{\hss#1\hss}}

\begin{document}

\title{Spectral Simplicity of Apparent Complexity, Part I:\\
The Nondiagonalizable Metadynamics of Prediction}

\author{Paul M. Riechers}
\email{pmriechers@ucdavis.edu}

\author{James P. Crutchfield}
\email{chaos@ucdavis.edu}

\affiliation{Complexity Sciences Center\\
Department of Physics\\
University of California at Davis\\
One Shields Avenue, Davis, CA 95616}

\date{\today}
\bibliographystyle{unsrt}

\begin{abstract}

Virtually all questions that one can ask about the behavioral and structural
complexity of a stochastic process reduce to a linear algebraic framing of a
time evolution governed by an appropriate hidden-Markov process generator. Each
type of question---correlation, predictability, predictive cost, observer
synchronization, and the like---induces a distinct generator class. Answers are
then functions of the class-appropriate transition dynamic. Unfortunately,
these dynamics are generically nonnormal, nondiagonalizable, singular, and so
on. Tractably analyzing these dynamics relies on adapting the recently
introduced \emph{meromorphic functional calculus}, which specifies the spectral
decomposition of functions of nondiagonalizable linear operators, even when the
function poles and zeros coincide with the operator's spectrum. Along the way,
we establish special properties of the projection operators that demonstrate
how they capture the organization of subprocesses within a complex system.
Circumventing the spurious infinities of alternative calculi, this
leads in the sequel, Part II, to the first closed-form expressions for
complexity measures, couched either in terms of the Drazin inverse
(negative-one power of a singular operator) or the eigenvalues and projection
operators of the appropriate transition dynamic.
\end{abstract}


\preprint{Santa Fe Institute Working Paper 17-05-XXX}
\preprint{arxiv.org:1705.XXXX [nlin.CD]}

\maketitle


\tableofcontents  

\setstretch{1.1}

\newcommand{\Abet}{\ProcessAlphabet}
\newcommand{\MS}{\MeasSymbol}
\newcommand{\ms}{\meassymbol}
\newcommand{\SSet}{\CausalStateSet}
\newcommand{\St}{\CausalState}
\newcommand{\st}{s}
\newcommand{\cs}{\causalstate}
\newcommand{\syncMSP}{{$\mathscr{S}$-MSP}}
\newcommand{\crypticMSP}{{$\PC$-MSP}}
\newcommand{\MxSt}{\AlternateState}
\newcommand{\MxSSet}{\AlternateStateSet_\pi}
\newcommand{\MxFSet}{\AlternateStateSet_\one}
\newcommand{\mxst}{\eta}
\newcommand{\mxstw}[1]{\mxst_{#1}} 		
\newcommand{\StartMS}{\bra{\delta_\pi}}
\newcommand{\opGen}{A}         
\newcommand{\matHMM}{T}     
\newcommand{\matMSP}{W}     
\newcommand{\TentT}{\varTheta}

\newcommand{\HWA}{\ket{\H(W^\Abet)}}
\newcommand{\Hmxst}{\ket{\H[\mxst]}}
\newcommand{\hsym}{\reflectbox{h}\text{h}}
\newcommand{\Redundancy}{\boldsymbol{R}}
\newcommand{\ACEphemeral}{\gamma_{\multimap}}
\newcommand{\hEphemeral}{h_{\multimap}}
\newcommand{\ACPersistent}{\gamma_{\rightsquigarrow}}
\newcommand{\hPersistent}{h_{\rightsquigarrow}}
\newcommand{\EEEphemeral}{\EE_{\multimap}}
\newcommand{\EEPersistent}{\EE_{\rightsquigarrow}}
\newcommand{\PU}{\reflectbox{$\mathcal{H}$}}

\newcommand{\kB}{k_\text{B}}
\newcommand{\corrbra}{\bra{\pi \overline{\Abet}}}
\newcommand{\corrket}{\ket{\Abet \one}}

\newcommand{\Cmatrix}{\mathcal{C}}   
\newcommand{\LWwoutZero}{\Lambda_W^{\setminus 0}}

\section{Introduction}

Complex systems---that is, many-body systems with strong interactions---are
usually observed through low-resolution feature detectors. The consequence is
that their hidden structure is, at best, only revealed over time. Since
individual observations cannot capture the full resolution of each degree of
freedom, let alone a sufficiently full set of them, measurement time series
often appear stochastic and non-Markovian, exhibiting long-range correlations.
Empirical challenges aside, restricting to the purely theoretical domain, even
finite systems can appear quite complicated. Despite admitting finite
descriptions, stochastic processes with sofic support, to take one example,
exhibit infinite-range dependencies among the chain of random variables they
generate \cite{Crut01a}. While such infinite-correlation processes are legion
in complex physical and biological systems, even approximately analyzing them
is generally appreciated as difficult, if not impossible. Generically, even
finite systems lead to uncountably infinite sets of predictive features
\cite{Marz17a}. These facts seem to put physical sciences' most basic
goal---prediction---out of reach.

We aim to show that this direct, but sobering conclusion is too bleak. Rather,
there is a collection of constructive methods that address hidden structure and
the challenges associated with predicting complex systems. This follows up on
our recent introduction of a functional calculus that uncovered new
relationships among supposedly different complexity measures~\cite{Crut13a} and
that demonstrated the need for a generalized spectral theory to answer such
questions~\cite{Riec16a}. Those efforts yielded elegant, closed-form solutions
for complexity measures that, when compared, offered insight into the overall
theory of complexity measures. Here, providing the necessary background for and
greatly expanding those results, we show that different questions regarding
correlation, predictability, and prediction each require their own analytical
structures, expressed as various kinds of hidden transition dynamic. The
resulting transition dynamic among hidden variables summarizes symmetry
breaking, synchronization, and information processing, for example. Each of
these metadynamics, though, is built up from the original given system.

The shift in perspective that allows the new level of tractability begins by
recognizing that---beyond their ability to \emph{generate} many sophisticated
processes of interest---hidden Markov models can be treated as exact
\emph{mathematical objects} when analyzing the processes they generate.
Crucially, and especially when addressing nonlinear processes, most questions
that we ask imply a linear transition dynamic over \emph{some} hidden state
space. Speaking simply, something happens, then it evolves linearly in time,
then we snapshot a selected characteristic. This broad type of sequential
questioning \emph{cascades}, in the sense that the influence of the initial
preparation cascades through state space as time evolves, affecting the final
measurement. Alternatively, other, complementary kinds of questioning involve
\emph{accumulating} such cascades. The linear algebra underlying either kind is
highlighted in Table \ref{table:SimpleLinAlgOfQs} in terms of an appropriate
discrete-time transition operator $T$ or a continuous-time generator $G$ of
time evolution.

\begin{table}
\begin{center}
\begin{tabular}{| l | c | c |}
\hline
\multicolumn{3}{|c|}{Linear Algebra Underlying Complexity} \\
\hline
Question type & Discrete time  & Continuous time  \\
\hline
Cascading
	& $\phantom{\Bigl( }\braket{ \cdot | T^L | \cdot} \phantom{\Bigr)}$
	&   $\braket{ \cdot | e^{t G} | \cdot}$ \\
\hline
Accumulating
	& $\phantom{\Bigl( } \braket{ \cdot | \left( \sum_L T^L \right) | \cdot} \phantom{\Bigr)}$
	&   $\braket{ \cdot | \left( \int e^{t G} \, dt \right) | \cdot}$ \\
\hline
\end{tabular}
\end{center}
\caption{Having identified the hidden linear dynamic, either a discrete-time
	operator $T$ or continuous-time operator $G$, quantitative questions tend
	to be either \emph{cascading} or \emph{accumulating} type. What changes
	between distinct questions are the dot products with the initial setup
	$\bra{\cdot}$ and the final observations $\ket{\cdot}$.
	}
\label{table:SimpleLinAlgOfQs}
\end{table}

In this way, deploying linear algebra to analyze complex systems turns on
identifying an appropriate hidden state space. And, in turn, the latter depends
on the genre of the question. Here, we focus on closed-form expressions for a
process' complexity measures. This determines what the internal system setup
$\bra{\cdot}$ and the final detection $\ket{\cdot}$ should be. We show that
complexity questions fall into three subgenres and, for each of these, we
identify the appropriate linear dynamic and closed-form expressions for several
of the key questions in each genre. See Table
\ref{table:SimpleQClassification}. The burden of the following is to explain
the table in detail. We return to a much-elaborated version at the end.

Associating observables $\ms \in \Abet$ with transitions between hidden states
$\st \in \SSet$, gives a hidden Markov model (HMM) with observation-labeled
transition matrices $\bigl\{ T^{(\ms)} : T^{(\ms)}_{i,j} = \Pr(\ms, \st_j |
\st_i ) \bigr\}_{\ms \in \Abet}$. They sum to the row-stochastic state-to-state
transition matrix $T = \sum_{\ms \in \Abet} T^{(\ms)}$. (The continuous-time
versions are similarly defined, which we do later on.) Adding measurement
symbols $\ms \in \Abet$ this way---to transitions---can be considered a model
of measurement itself \footnote{While we follow Shannon \cite{Shan48a} in this,
it differs from the more widely used state-labeled HMMs.}. The efficacy of our
choice will become clear.

It is important to note that HMMs, in continuous and discrete time, arise
broadly in the sciences, from quantum mechanics \cite{Moor97a,Clar15a},
statistical mechanics \cite{Penr70a}, and stochastic thermodynamics
\cite{Seif12a,Klag13a,Beck15a} to communication theory \cite{Shan48a,Cove06a},
information processing \cite{Rabi86a,Robe87,Rabi89a}, computer design
\cite{Rabi63a}, population and evolutionary dynamics \cite{Ewen04a,Nowa06a},
and economics. Thus, HMMs appear in the most fundamental physics and in the
most applied engineering and social sciences. The breadth suggests that the
thorough-going HMM analysis developed here is worth the required effort.

Since complex processes have highly structured, directional transition
dynamics---$T$ or $G$---we encounter the full richness of matrix algebra in
analyzing HMMs. We explain how analyzing complex systems \emph{induces} a
nondiagonalizable metadynamics, even if the original dynamic is diagonalizable
in its underlying state-space. Normal and diagonalizable restrictions, so
familiar in mathematical physics, simply fail us here.

The diversity of nondiagonalizable dynamics presents a technical challenge,
though. A new calculus for functions of nondiagonalizable operators---e.g.,
$T^L$ or $e^{tG}$---becomes a necessity if one's goal is an exact analysis of
complex processes. Moreover, complexity measures naively and easily lead one to
consider illegal operations. Taking the inverse of a singular operator is a
particularly central, useful, and fraught example. Fortunately, such illegal
operations can be skirted since the complexity measures only extract the excess
transient behavior of an infinitely complicated orbit space.

To explain how this arises---how certain modes of behavior, such as excess
transients, are selected as relevant, while others are
ignored---Ref.~\cite{Riec16a} recently developed a meromorphic functional
calculus for analyzing complex processes generated by HMMs. The following shows
that this leads to a general spectral theory of weighted directed graphs and
that, more specifically, the techniques can be applied to the challenges of
prediction. The results developed here greatly extend and (finally) explain
those announced in Ref.~\cite{Crut13a}. The latter introduced the basic methods
and results by narrowly focusing on closed-form expressions for several
measures of intrinsic computation, applying them to prototype complex systems.

The meromorphic functional calculus, summarized in detail later, concerns
functions of nondiagonalizable operators when poles (or zeros) of the function
of interest coincide with poles of the operator's resolvent---poles that appear
precisely at the eigenvalues of the transition dynamics. Pole--pole and
pole--zero interactions transform the complex-analysis residues within the
functional calculus. One notable result is that the negative-one power of a
singular operator exists in the meromorphic functional calculus. We derive its
form, note that it is the \emph{Drazin inverse}, and show how widely useful and
common it is.

For example, the following gives the first closed-form expressions for many
complexity measures in wide use---many of which turn out to be expressed most
concisely in terms of a Drazin inverse. Furthermore, spectral decomposition
gives insight into subprocesses of a complex system in terms of the projection
operators of the appropriate transition dynamic.

\begin{table}
\begin{center}
\begin{tabular}{| l | p{2.4cm} p{1.7cm} | p{2.45cm} |}
\hline
\multicolumn{4}{|c|}{Questions and Their Linear Dynamics} \\
\hline
\multicolumn{1}{|c|}{Genre} & \multicolumn{2}{c}{Measures} & \multicolumn{1}{|c|}{Hidden dynamic} \\
\hline
\multirow{2}{*}{Observation}
	& Correlations & $\gamma(L)$
	& \multirow{2}{*}{HMM matrix $T$} \\
	& Power spectra & $P(w)$ & \\
\hline
\multirow{2}{*}{Predict\emph{ability}}
	& Myopic entropy & $\hmu(L)$
	& HMM MSP \\
	& Excess entropy & $\EE$, $\EE(w)$ & matrix $W$ \\
\hline
\multirow{2}{*}{Predic\emph{tion}}
	& Causal & $\Cmu$, $\mathcal{H}^+ (L)$
	& \EM\ MSP \\
	& ~~~synchrony & $\mathbf{S}$, $S(w)$ & matrix $\mathcal{W}$ \\
\hline
\multirow{2}{*}{Generation}
	& State & $\Cmu(\mathcal{M})$,
	& Generator \\
	& ~~~synchrony & $\mathcal{H} (L)$, $\mathbf{S}^\prime$ & MSP matrix \\
\hline
\end{tabular}
\end{center}
\caption{Question genres (leftmost column) about process complexity listed with
	increasing sophistication. Each genre implies a different linear transition
	dynamic (rightmost column). \emph{Observational} questions concern the
	superficial, given dynamic. \emph{Predictability} questions are about the
	observation-induced dynamic over distributions; that is, over states used
	to generate the superficial dynamic. \emph{Prediction} questions address
	the dynamic over distributions over a process' causally-equivalent
	histories.  \emph{Generation} questions concern the dynamic over any
	nonunifilar presentation $\mathcal{M}$.
}
\label{table:SimpleQClassification}
\end{table}

To get started, sections \S \ref{sec:Processes} through \S \ref{sec:HMMs}
briefly review relevant background in stochastic processes, the HMMs that
generate them, and complexity measures. Several classes of HMMs are discussed in
\S \ref{sec:HMMs}. Mixed-state presentations (MSPs)---HMM generators of a
process that also track distributions induced by observation---are reviewed in
\S \ref{sec:MSPs}. They are key to complexity measures within an
information-theoretic framing. Section \S \ref{sec:FindingLinearity} then shows
how each complexity measure reduces to the linear algebra of an appropriate HMM
adapted to the question genre.

To make progress at this point, we summarize the meromorphic functional
calculus in \S \ref{sec:SpectralTheory}. Several of its mathematical
implications are discussed in relation to projection operators in \S
\ref{sec:ProjOpsForStochasticMatrices} and a spectral weighted directed graph
theory is presented in \S \ref{sec:SpectraByInSpection}.

With this all set out, the sequel Part II finally derives the promised
closed-form complexities of a process and outlines common simplifications for
special cases. Leveraging the functional calculus, it introduces a novel
extension---the complexity measure frequency spectrum and shows how to calculate
it in closed form. It provides a suite of examples to ground the theoretical
developments and works through in-depth a pedagogical example.

\section{Structured Processes and their Complexities} 
\label{sec:Processes}

We first describe a system of interest in terms of its observed behavior,
following the approach of computational mechanics, as reviewed in
Ref.~\cite{Crut12a}. Again, a \emph{process} is the collection of behaviors
that the system produces and their probabilities of occurring. A process's
behaviors are described via a bi-infinite chain of random variables, denoted by
capital letters $\ldots \, \MS_{t-2} \, \MS_{t-1} \, \MS_{t} \, \MS_{t+1} \,
\MS_{t+2} \ldots$.  A realization is indicated by lowercase letters $\ldots \,
\ms_{t-2} \, \ms_{t-1} \, \ms_{t} \, \ms_{t+1} \, \ms_{t+2} \ldots$. We assume
values $\ms_{t}$ belong to a discrete alphabet $\MeasAlphabet$. We work with
blocks $\MS_{t:t^\prime}$, where the first index is inclusive and the second
exclusive: $\MS_{t:t^\prime} = \MS_t \ldots \MS_{t^\prime-1}$. Block
realizations $\ms_{t:t^\prime}$ we often refer to as \emph{words} $w$. At each
time $t$, we can speak of the \emph{past} $\MS_{-\infty:t} = \ldots \MS_{t-2}
\MS_{t-1}$ and the \emph{future} $\MS_{t:\infty} = \MS_{t} \MS_{t+1}
\ldots$.

A process's probabilistic specification is a density over these chains:
$\mathbb{P}(\MS_{-\infty:\infty})$. Practically, we work with finite blocks and
their probability distributions $\Prob(\MS_{t:t^\prime})$. To simplify the
development, we primarily analyze stationary, ergodic processes: those for
which $\Prob(\MS_{t:t+L}) = \Prob(\MS_{0:L})$ for all $t \in \mathbb{Z}$, $L
\in \mathbb{Z}^+$, and all realizations. In such cases, we only need to
consider a process's length-$L$ \emph{word distributions} $\Prob(\MS_{0:L})$.

\subsection{Directly observable organization}

A common first step to understand how processes express themselves is to
analyze correlations among observables. Pairwise correlation in a sequence of
observables is often summarized by the \emph{autocorrelation function}:
\begin{align*}
\gamma(L) & = \left\langle \overline{\MS}_t \MS_{t+L} \right\rangle_t 
  ~,
\end{align*}
where the bar above $\MS_t$ denotes its complex conjugate, and the angled
brackets denote an average over all times $t \in \mathbb{Z}$. Alternatively,
structure in a stochastic process is often summarized by the \emph{power
spectral density}, also referred to more simply as the \emph{power spectrum}:
\begin{align*}
\Psp = 
\lim_{N \to \infty} 
  \frac{1}{N} \, 
      \left\langle \, 
        \biggl| \sum_{L = 1}^N X_L e^{- i \omega L}
        \biggr|^2
      \right\rangle
  ~,
\end{align*}
where 
$\omega \in \mathbb{R}$ is the angular frequency~\cite{Stoi05}. Though a basic
fact, it is not always sufficiently emphasized in applications that power
spectra capture only pairwise correlation. Indeed, it is straightforward to
show that the power spectrum $\Psp$ is the windowed Fourier transform of
the autocorrelation function $\gamma(L)$. That is, power spectra describe how
pairwise correlations are distributed across frequencies. Power spectra are
common in signal processing, both in technological settings and physical
experiments \cite{Hamm97a}. As a physical example, diffraction patterns are the
power spectra of a sequence of structure factors \cite{Wool97a}.

To monitor transport properties in near-equilibrium thermodynamic systems, the
Green--Kubo coefficients are another important example measure of observable
organization, but are rather more application-specific~\cite{Green54, Zwan65}.
These coefficients reflect the idea that dissipation depends on correlation
structure. They usually appear in the form of integrating the autocorrelation
of \emph{derivatives} of observables. A change of observables, however, turns
this into an integration of a standard autocorrelation function. Green--Kubo
transport coefficients then involve the limit $\lim_{\omega \to 0} \Psp$ for
the process of appropriate observables.

One theme in the following is that, though widely used, correlation functions
and power spectra give an impoverished view of a process's structural
complexity, since they only consider ensemble averages over pairwise events.
Moreover, creating a list of higher-order correlations is an impractical way to
summarize complexity, as seen in the \emph{connected correlation functions} of
statistical mechanics \cite{Binn92a}.

\subsection{Intrinsic predictability} 
\label{sec:Predictability}

Information measures, in contrast, can involve all orders of correlation and
thus help to go beyond pairwise correlation in understanding, for example,  how
a process' past behavior affects predicting it at later times. Information
theory, as developed for general complex processes \cite{Crut01a}, provides a
suite of quantities that capture prediction properties using variants of
Shannon's entropy $\H[\cdot]$ and mutual information $\I[\, \cdot \, ; \cdot \,
]$ \cite{Cove06a} applied to sequences. Each measure answers a specific
question about a process' predictability. For example:
\begin{itemize}
\setlength{\itemsep}{-4pt}
\setlength{\topsep}{-6pt}
\setlength{\parsep}{-6pt}
\setlength\itemindent{-15pt}
\item How much information is contained in the words generated?
	The \emph{block entropy} \cite{Crut01a}:\\
\begin{align*}
\H(L) = - \sum_{w \in \Abet^L} \Prob(w) \log_2 \Prob(w)
  ~.
\end{align*}
\item How random is a process?
	Its \emph{entropy rate} \cite{Kolm59}:\\
\begin{align*}
\hmu = \lim_{L \to \infty} \H(L)/L
  ~.
\end{align*}
\item How is the irreducible randomness $\hmu$ approached?
	Via the \emph{myopic entropy rates} \cite{Crut01a}:\\
\begin{align*}
\hmu(L) = \H[\MS_0 | \MS_{1-L} \ldots \MS_{-1}]
  ~.
\end{align*}
\item How much of the future can be predicted?
	Its \emph{excess entropy} \cite{Crut01a}:\\
\begin{align*}
\EE = \I[\MS_{-\infty:0} ; \MS_{0:\infty} ]
  ~.
\end{align*}
\item How much information must be extracted to know its predictability and
	so see its intrinsic randomness $\hmu$? Its \emph{transient information}
	\cite{Crut01a}:
\begin{align*}
\TI = \sum_{L=0}^\infty \left[ \EE + \hmu L - \H(L) \right]
  ~.
\end{align*}
\end{itemize}

The spectral approach, our subject, naturally leads to allied, but new
information measures. To give a sense, later we introduce the \emph{excess
entropy spectrum} $\EEsp$. It completely, yet concisely, summarizes the
structure of myopic entropy reduction, in a way similar to how the power
spectrum completely describes autocorrelation. However, while the power
spectrum summarizes only pairwise linear correlation, the excess entropy
spectrum captures all orders of nonlinear dependency between random variables,
making it an incisive probe of hidden structure.

Before leaving the measures related to predictability, we must also point out
that they have important refinements---measures that lend a particularly
useful, even functional, interpretation. These include the bound, ephemeral,
elusive, and related informations \cite{Jame11a,Ara14a}. Though amenable to the
spectral methods of the following, we leave their discussion for another venue.
Fortunately, their spectral development is straightforward, but would take us
beyond the minimum necessary presentation to make good on the overall
discussion of spectral decomposition.

\subsection{Prediction overhead}
\label{sec:PredictiveBurden}

Process predictability measures, as just enumerated, certainly say much about a
process' intrinsic information processing. They leave open, though, the
question of the structural complexity associated with \emph{implementing}
prediction. This challenge entails a complementary set of measures that
directly address the inherent complexity of actually \emph{predicting} what is
predict\emph{able}. For that matter, how cryptic is a process?

Computational mechanics describes optimal prediction via a process' hidden,
effective or causal states and transitions, as summarized by the process's
\eM~\cite{Crut12a}. A \emph{causal state} $\cs \in \SSet^{\forward}$ is an
equivalence class of histories $\MS_{-\infty:0}$ that all yield the same
probability distribution over observable futures $\MS_{0:\infty}$. Therefore,
knowing a process's current causal state---that $\St_0^{\forward} = \cs$,
say---is sufficient for optimal prediction.

Computational mechanics provides an additional suite of quantities that capture
the overhead of prediction, again using variants of Shannon's entropy and
mutual information applied to the \eM. Each also answers a specific question
about an observer's burden of prediction. For example:
\begin{itemize}
\setlength{\itemsep}{-4pt}
\setlength{\topsep}{-6pt}
\setlength{\parsep}{-6pt}
\setlength\itemindent{-15pt}
\item How much historical information must be stored for optimal prediction?
	The \emph{statistical complexity} \cite{Crut88a}:
\begin{align*}
\Cmu = \H[\St_0^{\forward}]
   ~.
\end{align*}
\item How unpredictable is a causal state upon observing a process for
duration $L$? The \emph{myopic causal-state uncertainty} \cite{Crut01a}:
\begin{align*}
\mathcal{H}^{\forward} (L) = \H[\CausalState_0^{\forward} | \MS_{-L} \ldots \MS_{-1}]
  ~.
\end{align*}
\item How much information must an observer extract to synchronize to---that is, to know with certainty---the causal state? The optimal predictor's \emph{synchronization information} \cite{Crut01a}:
\begin{align*}
\SI = \sum_{L=0}^\infty \mathcal{H}^{\forward}  (L)
  ~. 
\end{align*}
\end{itemize}

Paralleling the purely informational suite of the previous section, we later
introduce the \emph{optimal synchronization spectrum} $\SIsp$. It completely
and concisely summarizes the frequency distribution of state-uncertainty
reduction, similar to how the power spectrum $\Psp$ completely describes
autocorrelation and the excess entropy spectrum $\EEsp$ the myopic entropy
reduction. Helpfully, the above optimal prediction measures can be found from
the optimal synchronization spectrum.

The structural complexities monitor an observer's burden in optimally
predicting a process. And so, they have practical relevance when an intelligent
artificial or biological agent must take advantage of a structured stochastic
environment---e.g., a Maxwellian Demon taking advantage of correlated
environmental fluctuations \cite{Boyd16d}, prey avoiding easy prediction, or
profiting from stock market volatility, come to mind.

Prediction has many natural generalizations. For example, since optimal
prediction often requires infinite resources, suboptimal prediction is of
practical interest. Fortunately, there are principled ways to investigate the
tradeoffs between predictive accuracy and computational burden \cite{Stil07b,
Creu09, Marz16_RateDistortion,Marz17a}. As another example, optimal prediction
in the presence of noisy or irregular observations can be investigated with a
properly generalized framework; see Ref. \cite{Elli13a}. Blending the existing
tools, resource-limited prediction under such observational constraints can
also be investigated. In all of these settings, information measures similar to
those listed above are key to understanding and quantifying the tradeoffs
arising in prediction.

Having highlighted the difference between prediction and predictability, we can
appreciate that some processes hide more internal information---are more
cryptic---than others. It turns out, this can be quantified. The
\emph{crypticity} $\PC = \Cmu - \EE$ is the difference between the a process's
stored information $\Cmu$ and the mutual information $\EE$ shared between past
and future observables~\cite{Maho11a}. Operationally, crypticity contrasts
predictable information content $\EE$ with an observer's minimal stored-memory
overhead $\Cmu$ required to \emph{make} predictions. To predict what is
predictable, therefore, an optimal predictor must account for a process's
crypticity.

\subsection{Generative complexities}

\newcommand{\ASt}{\AlternateState}
\newcommand{\ASet}{\AlternateStateSet}
\renewcommand{\ast}{\alternatestate}

How does a physical system produce its output process? This depends on many
details. Some systems employ vast internal mechanistic redundancy, while others
under constraints have optimized internal resources down to a minimally
necessary generative structure. Different pressures give rise to different
kinds of optimality. For example, minimal state-entropy generators turn out to
be distinct from minimal state-set generators \cite{Crut92c,Uppe97a,Lohr09b}.
The challenge then is to develop ways to monitor differences in generative
mechanism.

Any generative model~\cite{Crut10a, Crut01a} $\mathcal{M}$ with state-set
$\AlternateStateSet$ has a statistical complexity (state entropy):
$C(\mathcal{M}) = \H[\ASt]$. Consider the corresponding \emph{myopic
state-uncertainty} given $L$ sequential observations:
\begin{align*}
\mathcal{H} (L) = \H[\ASt_0 |  \MS_{-L:0} ]
  ~, 
\end{align*} 
And so:
\begin{align*}
\mathcal{H} (0) = C(\mathcal{M}) 
~.
\end{align*}
We also have the asymptotic uncertainty $\mathcal{H} \equiv \lim_{L \to \infty}
\mathcal{H} (L)$.
Related, there is the \emph{excess synchronization information}:
\begin{align*} 
\SI' = \sum_{L=0}^\infty \bigl[ \mathcal{H} (L) - \mathcal{H} \bigr] ~.
\end{align*}
Such quantities are relevant even when an observer never fully synchronizes to
a generative state; i.e., even when $\mathcal{H} > 0$. Finite-state \eMs\
always synchronize \cite{Trav10a,Trav10b} and so their $\mathcal{H}$ vanishes.

Since many different mechanisms can generate a given process, we need useful
bounds on the statistical complexity of possible process generators. For
example, the minimal \emph{generative complexity} $C_\text{g} =
\min_{\{\AlternateStateSet\}} C(\mathcal{M})$ is the minimal state-information
a physical system must store to generate its future \cite{Lohr09b}. The
predictability and the statistical complexities bound each other:
\begin{align*}
\EE \leq C_\text{g} \leq \Cmu
  ~.
\end{align*} 
That is, the \emph{predictable} future information $\EE$ is less than or equal
to the information $C_\text{g}$ necessary to \emph{produce} the future which,
in turn, is less than or equal to the information $\Cmu$ necessary to
\emph{predict} the future \cite{Crut92c,Uppe97a,Crut01a,Lohr09b}. Such
relationships have been explored even for quantum generators of (classical)
stochastic processes \cite[and references therein]{Maho15a}.

\section{Hidden Markov Models}
\label{sec:HMMs}

Up to this point, the development focused on introducing and interpreting
various information and complexity measures. It was not constructive in that
there was no specification of how to calculate these quantities for a given
process. To do so requires models or, in the vernacular, a \emph{presentation}
of a process. Fortunately, a common mathematical representation describes a
wide class of process generators: the edge-labeled \emph{hidden Markov models}
(HMMs), also known as a Mealy HMMs~\cite{Crut10a} \footnote{Contrast this with
the class-equivalent state-labeled HMMs, also known as Moore HMMs~\cite{Bala93a,
Uppe97a, Vanluyten08}. In automata theory, a finite-state HMM is called a
\emph{probabilistic nondeterministic finite automaton}~\cite{Sips12}.
Information theory \cite{Cove06a} refers to them as finite-state
\emph{information sources}. And, stochastic process theory defines them as
\emph{functions of a Markov chain}~\cite{Blac57b,Birc61a, Ash65,Trav14}}. Using
these as our preferred presentations, we will first classify them and then
describe how to calculate the information measures of the processes they
generate.

\begin{Def}
\label{def:HMM}
A \emph{finite-state, edge-labeled hidden Markov model}
$\mathcal{M} = \bigl\{ \ASet, \Abet, \{ T^{(\ms)} \}_{\ms \in \Abet}, \mxst_0
\bigr\}$
consists of:
\begin{itemize}
\setlength{\itemsep}{-4pt}
\setlength{\topsep}{-6pt}
\setlength{\parsep}{-6pt}
\setlength\itemindent{-15pt}
\item A finite set of hidden states
	$\AlternateStateSet = \left\{ \ast_{1}, \ldots ,\ast_{M} \right\}$.
	$\ASt_t$ is the random variable for the hidden state at time $t$.
\item A finite output alphabet $\Abet$.
\item A set of $M \times M$ symbol-labeled transition matrices
	$\bigl\{ T^{(\ms)} \bigr\}_{\ms \in \Abet}$, where $T^{(\ms)}_{i,j} =
	\Pr(\ms, \ast_{j} | \ast_{i})$ is the probability of transitioning from
	state $\ast_{i}$ to state $\ast_{j}$ and emitting symbol $\ms$. The
	corresponding overall state-to-state transition matrix is the
	row-stochastic matrix $T = \sum_{\ms \in \Abet} T^{(\ms)}$.
\item An initial distribution over hidden states:
	$\mxst_0 = \bigl( \Pr(\ASt_0 = \ast_1), \Pr(\ASt_0 = \ast_2), ...,
	\Pr(\ASt_0 = \ast_M) \bigr)$.
\end{itemize}
\end{Def}

The dynamics of such finite-state models are governed by transition matrices
amenable to the linear algebra of vector spaces. As a result, bra-ket notation
is useful \cite{Dira39a}. Bras $\bra{\cdot}$ are row vectors and kets
$\ket{\cdot}$ are column vectors. One benefit of the notation is immediately
recognizing mathematical object type. For example, on the one hand, any
expression that forms a closed bra-ket pair---either $\braket{\cdot | \cdot}$
or $\bra{\cdot} \cdot \ket{\cdot}$---is a scalar quantity and commutes as a
unit with anything. On the other hand, when useful, an expression of the
ket-bra form $\ket{\cdot} \bra{\cdot}$ can be interpreted as a matrix.

$T$'s row-stochasticity means that each of its rows sum to unity. Introducing
$\ket{\one}$ as the column vector of all 1s, this can be restated as:
\begin{align*}
T \ket{\one} = \ket{\one} ~.
\end{align*}
This is readily recognized as an eigenequation: $T \ket{\eta} = \lambda
\ket{\eta}$. That is, the all-ones vector $\ket{\one}$ is always a right
eigenvector of $T$ associated with the eigenvalue $\lambda$ of unity.

When the internal Markov transition matrix $T$ is irreducible, the
Perron-Frobenius theorem guarantees that there is a unique asymptotic state
distribution $\pi$ determined by:
\begin{align*}
\bra{\pi} T = \bra{\pi} ~,
\end{align*}
with the further condition that $\pi$ is normalized in probability:
$\braket{\pi | \one} = 1$. This again is recognized as an eigenequation: the
asymptotic distribution $\pi$ over the hidden states is $T$'s left eigenvector
associated with the eigenvalue of unity.

To describe a stationary process, as done often in the following, the initial
hidden-state distribution $\mxst_0$ is set to the asymptotic one: $\mxst_0 =
\pi$. The resulting process generated is then stationary. Choosing an
alternative $\mxst_0$ is useful in many contexts, but yields a nonstationary
process. We avoid this for now for simplicity.

An HMM $\mathcal{M}$ describes a process' behaviors as a formal \emph{language}
$\mathcal{L} \subseteq \bigcup_{\ell = 1}^\infty \mathcal{A}^\ell$ of allowed
realizations. Moreover, $\mathcal{M}$ succinctly describes a process's word
distribution $\Pr(w)$ over all words $w \in \mathcal{L}$. (Appropriately,
$\mathcal{M}$ also assigns zero probability to words outside of the process'
language: $\Pr(w) = 0$ for all $w \in \mathcal{L}^\text{c}$, $\mathcal{L}$'s
complement.) Specifically, the stationary probability of observing a particular
length-$L$ word $w = \ms_0 \ms_1 \ldots \ms_{L-1}$ is given by:
\begin{align}
\Pr(w) =  \bra{\pi} T^{(w)} \ket{\one} ~,
\label{eq:WordDistHMM}
\end{align}
where $T^{(w)} \equiv T^{(\ms_0)} T^{(\ms_1)} \dotsm T^{(\ms_{L-1})}$.

More generally, given a nonstationary state distribution $\eta$, the subsequent
probability of a word is:
\begin{align}
\Pr(\MS_{t:t+L} = w | \ASt_t \sim \mxst ) =  \bra{\mxst} T^{(w)} \ket{\one} ~,
\label{eq:WordDistHMMnoneq}
\end{align}
where $\ASt_t \sim \mxst$ means that the random variable $\ASt_t$ is
distributed as $\mxst$ \cite{Cove06a}. This conditional word probability is
used often since, for example, most observations induce a nonstationary
distribution over hidden states. Tracking such observation-induced
distributions is the role of a related model class---the mixed-state
presentation, introduced shortly. To get there, we must first introduce
several, prerequisite HMM classes. See Fig. \ref{fig:ExampleHMMs}. The general
HMM just discussed is shown in Fig. \ref{fig:NonunifilarHMM}.

\subsection{Unifilar HMMs}

An important class of HMMs consists of those that are unifilar.
\emph{Unifilarity} guarantees that, given a start state and a sequence of
observations, there is a unique path through the internal states~\cite{Ash65}.
This, in turn, allows one to directly translate properties of the internal
Markov chain into properties of the observed behavior generated from the
sequence of edges traversed. Unifilar HMMs are a process' optimal predictors
\cite{Shal98a}.

In contrast, general---that is, nonunifilar---HMMs have an exponentially
growing number of possible state paths as a function of observed word length.
Thus, nonunifilar process presentations break most all quantitative connections
between internal dynamics and observations, rendering them markedly less useful
process presentations. While they can be used to generate realizations of a
given process, they cannot be used to predict a process. Unifilarity is
required.

\begin{Def}
\label{def:uHMM}
A \emph{finite-state, edge-labeled, unifilar HMM} (uHMM) \footnote{Automata
theory would refer to a uHMM as a \emph{probabilistic deterministic finite
automaton} \cite{Sips12}. The awkward terminology does not recommend itself.}
is a finite-state, edge-labeled HMM with the following property:
\begin{itemize}
\item \emph{Unifilarity}: For each state $\ast \in \ASet$ and each symbol
	$\ms \in \Abet$ there is at most one outgoing edge from state $\ast$ that
	emits symbol $\ms$.
\end{itemize}
\end{Def}

An example is shown in Fig. \ref{fig:UnifilarHMM}.

\subsection{Minimal unifilar HMMs}

Minimal models are not only convenient to use, but very often allow for
determining essential informational properties, such as a process' memory
$\Cmu$. A process' minimal state-entropy uHMM is the same as its minimal-state
uHMM. And, the latter turns out to be the process' \emph{\eM} in computational
mechanics~\cite{Crut12a}. Computational mechanics shows how to calculate a
process' \eM\ from the process' conditional word distributions. Specifically,
\eM\ states, the process' \emph{causal states} $\cs \in \SSet$, are equivalence
classes of histories that yield the same predictions for the future.
Explicitly, two histories $\past$ and $\pastprime$ map to the same causal state
$\epsilon(\past) = \epsilon(\pastprime) = \cs$ if and only if $\Pr(\Future |
\past) = \Pr(\Future | \pastprime)$. Thus, each causal state comes with a
prediction of the future $\Pr(\Future | \cs)$---its \emph{future morph}. In
short, a process' \eM\ is its minimal size, optimal predictor.

Converting a given uHMM to its corresponding \eM\ employs probabilistic
variants of well-known state-minimization algorithms in automata theory
\cite{Hopc79}. One can also verify that a given uHMM is minimal by checking
that all its states are probabilistically distinct \cite{Trav10a,Trav10b}.

\begin{Def}
\label{def:ProbDistinctStates}
A uHMM's states are \emph{probabilistically distinct} if for each pair of
distinct states $\ast_k, \ast_j \in \ASet$ there exists some finite word $w =
\ms_0 \ms_1 \ldots \ms_{L-1}$ such that:
\begin{align*}
\Pr(\Future = w | \ASt = \ast_k) \neq \Pr(\Future = w | \ASt = \ast_j)
  ~.
\end{align*}
\end{Def}
If this is the case, then the process' uHMM is its \eM.

An example is shown in Fig. \ref{fig:eM}.

\subsection{Finitary stochastic process hierarchy}

The finite-state presentations in these classes form a hierarchy in terms of
the processes they can finitely generate \cite{Crut92c}: Processes(\eMs) $=$
Processes(uHMMs) $\subset$ Processes(HMMs). That is, finite HMMs generate a
strictly larger class of stochastic processes than finite uHMMs. The class of
processes generated by finite uHMMs, though, is the same as generated by finite
\eMs.

\subsection{Continuous-time HMMs}
\label{sec:ContTimeHMMs}

Though we concentrate on discrete-time processes, many of the process
classifications, properties, and calculational methods carry over easily to
continuous time. In this setting transition \emph{rates} are more
appropriate than transition probabilities. Continuous-time HMMs can often be
obtained as a discrete-time limit $\Delta t \to 0$ of an edge-labeled HMM whose
edges operate for a time $\Delta t$. The most natural continuous-time HMM
presentation, though, has a continuous-time generator $G$ of time evolution
over hidden states, with observables emitted as deterministic \emph{functions
of an internal Markov chain}: $f: \SSet \to \Abet$.

Respecting the continuous-time analogue of probability conservation, each row
of $G$ sums to zero. Over a finite time interval $t$, marginalizing over all
possible observations, the row-stochastic state-to-state transition dynamic is:
\begin{align*}
T_{t_0 \to t_0 + t} = e^{t G} ~.
\end{align*}
The generated process, a function of the internal continuous-time Markov chain,
can also be specified by a set of transition matrices. For this purpose we
introduce the continuous-time \emph{observation matrices}:
\begin{align*}
\Gamma_x = \sum_{\ast \in \ASet}
  \delta_{x, f(\ast)} \ket{\delta_\ast} \bra{\delta_\ast}
  ~,
\end{align*} 
where $\delta_{x, f(\ast)}$ is a Kronecker delta, $\ket{\delta_\ast}$ the
column vector of all zeros except for a one at the position for state $\ast$,
and $\bra{\delta_\ast}$ its transpose $\bigl( \ket{\delta_\ast} \bigr)^\top$.
These ``projectors'' sum to the identity: $\sum_{x \in \Abet} \Gamma_x = I$.

\begin{figure*}
\centering
\subfloat[Nonunifilar HMM]{\label{fig:NonunifilarHMM}
\includegraphics[width=0.18\textwidth]{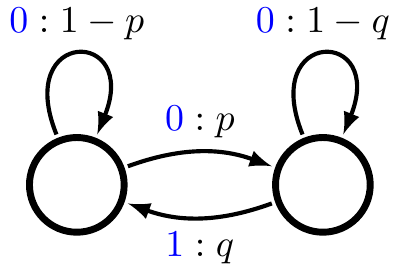} }
\subfloat[Unifilar HMM]{\label{fig:UnifilarHMM}
\includegraphics[width=0.29\textwidth]{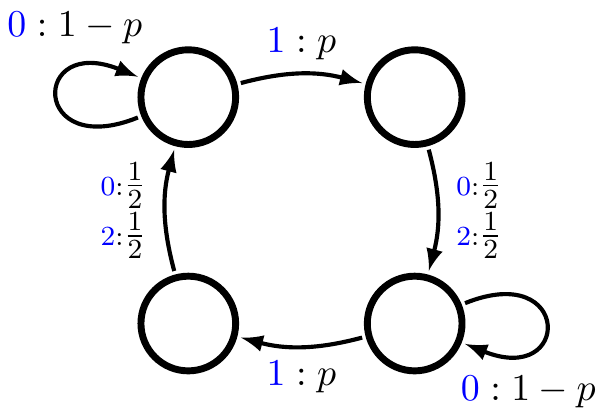} }
\subfloat[\eM]{\label{fig:eM}
\includegraphics[width=0.18\textwidth]{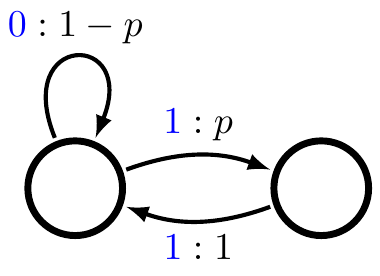} }
\subfloat[Continuous-time function of a Markov chain]{\label{fig:ContTimeHMM}
\includegraphics[width=0.29\textwidth]{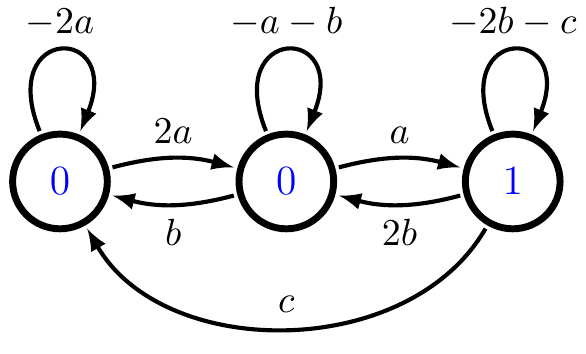} }
\caption{Example processes generated by the finite HMM classes, depicted
	by their state-transition diagrams: For any setting of the transition
	probabilities $p, q \in (0, \, 1)$ and transition rates $a, b, c \in (0, \,
	\infty)$, each HMM generates an observable stochastic process over its
	alphabet $\Abet \subset \{ {\color{blue}0}, {\color{blue}1},
	{\color{blue}2} \}$---the latent states themselves are not directly
	observable from the output process and so are ``hidden''.  
(a) Simple nonunifilar source: Two transitions leaving from the same state
	generate the same output symbol.
(b) Nonminimal unifilar HMM.
(c) \EM: Minimal unifilar HMM for the stochastic process generated.
(d) Generator of a continuous-time stochastic process.  
}
\label{fig:ExampleHMMs}
\end{figure*}

An example is shown in Fig. \ref{fig:ContTimeHMM}.

\section{Mixed-State Presentations}
\label{sec:MSPs}

A given process can be generated by nonunifilar, unifilar, and \eM\ HMM
presentations. Within either the unifilar or nonunifilar HMM classes, there can
be an unbounded number of presentations that generate the process. A process'
\eM\ is unique, however.  

This flexibility suggests that we can create a HMM process generator to answer
more refined questions than information generation ($\hmu$) and memory
($\Cmu$) calculated from the \eM. To this end, we introduce the \emph{mixed-state presentation} (MSP). An
MSP tracks important supplementary information \emph{in} the hidden states and,
through well-crafted dynamics, \emph{over} the hidden states.  In particular,
an MSP generates a process while tracking the observation-induced distribution
over the states of an alternative process generator. Here, we review only that
subset of mixed-state theory required by the following.

Consider a HMM presentation $\mathcal{M} = \bigl( \ASet, \Abet, \{ T^{(x)}
\}_{x \in \Abet}, \pi \bigr)$ of some process in statistical equilibrium. A
mixed state $\mxst$ can be any state distribution over $\ASet$, but the
uncountable set of points in the most general state-distribution simplex is
infinitely more than needed to calculate many complexity measures. How to
monitor the way in which an observer comes to know the HMM state as it sees
successive symbols from the process? This is the problem of
\emph{observer-state synchronization}. To analyze this evolution of the
observer's knowledge, we use the set $\MxSSet$ of mixed states that are induced
by all allowed words $w \in \mathcal{L}$ from initial mixed state $\mxst_0= \pi$:
\begin{align*}
\MxSSet = 
  \bigcup_{w \in \mathcal{L}}
  \frac{\bra{\pi} T^{(w)}}{\bra{\pi} T^{(w)} \ket{\one}}
  ~.
\end{align*}
The cardinality of $\MxSSet$ is finite when there are only a finite number of
distinct probability distributions over $\mathcal{M}$'s states that can be
induced by observed sequences, if starting from the stationary distribution
$\pi$.

If $w$ is the first (in lexicographic order) word that induces a particular distribution over $\ASet$,
then we denote this distribution as $\mxst^w$. For example, if the two words
$010$ and $110110$ both induce the same distribution $\eta$ over $\ASet$ and no
word shorter than $010$ induces that distribution, then the mixed state is
denoted $\mxst^{010}$. It corresponds to the distribution:
\begin{align*}
\bra{\mxst^{010}} & =
  \frac{\bra{\pi} T^{(0)} T^{(1)} T^{(0)}}
  {\bra{\pi} T^{(0)} T^{(1)} T^{(0)} \ket{\one} }
  ~.
\end{align*}

Since a given observed symbol induces a unique updated distribution from a
previous distribution, the dynamic over mixed states is unifilar. Transition
probabilities among mixed states can be obtained via
Eq.~\eqref{eq:WordDistHMMnoneq}. So, if:
\begin{align*}
\braket{\mxst | T^{(\ms)} | \one } > 0
\end{align*}
and:
\begin{align*}
\bra{\mxst' } =
  \frac{\bra{\mxst} T^{(\ms)}}{\braket{\mxst | T^{(\ms)} | \one }}
  ~,
\end{align*}
then:
\begin{align*}
\Pr(\mxst' , \ms | \mxst) & = \Pr(\ms | \mxst) \\
  & = \braket{\mxst | T^{(\ms)} | \one }
  ~.
\end{align*}
These transition probabilities over the mixed states in $\MxSSet$ are the
matrix elements for the observation-labeled transition matrices $\{ W^{(\ms)}
\}_{\ms \in \Abet}$ of $\mathcal{M}$'s \emph{synchronizing MSP} (\syncMSP):
\begin{align*}
\mathscr{S}\text{-MSP}(\mathcal{M}) 
= \bigl( \MxSSet, \Abet, \{ W^{(\ms)} \}_{\ms \in \Abet}, \delta_\pi \bigr)~,
\end{align*}
where $\delta_\pi$ is the distribution over $\MxSSet$ peaked at the unique
start-(mixed)-state $\pi$. The row-stochastic net mixed-state-to-state
transition matrix of \syncMSP($\mathcal{M}$) is $W = \sum_{\ms \in \Abet}
W^{(\ms)}$. If irreducible, then there is a unique stationary probability
distribution $\bra{\pi_W}$ over \syncMSP($\mathcal{M}$)'s states obtained by
solving $\bra{\pi_W} = \bra{\pi_W} W$. We use $\MxSt_t$ to denote the random
variable for the MSP's state at time $t$.

More generally, we must consider a mixed-state dynamic that starts from a
nonpeaked distribution over the hidden-state distribution simplex. This may be
counterintuitive, since a distribution over distributions should correspond to
a single distribution.  However, general MSP theory with a nonpeaked starting
distribution over the simplex allows us to consider a weighted average of
behaviors originating from disparate histories. And, this is distinct from considering the behavior originating from a weighted average of histories.
This more general MSP formalism arises in the closed-form solutions for more
sophisticated complexity measures, such as the bound information. This
appears in a sequel.

With this brief overview of mixed states, we can now turn to use them.
Section \S~\ref{sec:FindingLinearity} shows that tracking distributions over
the states of another generator makes the MSP an ideal algebraic object for
closed-form complexity expressions involving conditional entropies---measures
that require conditional probabilities. Sections \S~\ref{sec:Predictability}
and \S~\ref{sec:PredictiveBurden} showed that many of the complexity measures
for predictability and predictive burden are indeed framed as conditional
entropies. And so, MSPs are central to their closed-form expressions.

Historically, mixed states were already implicit in Ref.~\cite{Blac57b},
introduced in their modern form by Ref.~\cite{Crut92c,Uppe97a}, and have been
used recently; e.g., in Refs. \cite{Crut08a,Crut08b}. Most of these efforts,
however, used mixed-states in the specific context of the \emph{synchronizing
MSP} (\syncMSP). A greatly extended development of mixed-state dynamics appears
in Ref.~\cite{Elli13a}. Different information-theoretic questions require
different mixed-state dynamics, each of which is a unifilar presentation.
Employing the mathematical methods developed here, we find that desired
closed-form solutions are often simple functions of the transition dynamic of
an appropriate MSP. The spectral character of the relevant MSP controls the
behavior of information-theoretic quantities.

Finally, we emphasize that similar linear algebraic constructions---where
hidden states track relevant information---that are nevertheless \emph{not}
MSPs are just as important for answering a different set of questions about a
process. Since the other constructions are not directly about predictability
and prediction, we report on these findings elsewhere.

\section{Identifying the Hidden Linear Dynamic}
\label{sec:FindingLinearity}

We are now in a position to identify the hidden linear dynamic appropriate to
many of the questions that arise in complex systems---their observation,
predictability, prediction, and generation, as outlined in Table
\ref{table:SimpleQClassification}. In part, this section addresses a very
practical need for specific calculations. In part, it also lays the foundations
for further generalizations, to be discussed at the end. Identifying the linear
dynamic means identifying the linear operator $\opGen$ such that a question of
interest can be reformulated as either being of the cascading form
$\braket{\cdot | \opGen^n | \cdot}$ or as an accumulation of such cascading
events via $\braket{ \cdot | \left( \sum_n \opGen^n \right) | \cdot}$; recall
Table \ref{table:SimpleLinAlgOfQs}. Helpfully, many well-known questions of
complexity can be mapped to these archetypal forms. And so, we now proceed to
uncover the hidden linear dynamics of the cascading questions approximately in
the order they were introduced in \S~\ref{sec:Processes}.

\subsection{Simple complexity from any presentation}

For observable correlation, any HMM transition operator will do as the linear
dynamic. We simply observe, let time (or space) evolve forward, and observe
again. Let's be concrete.

Recall the familiar autocorrelation function. For a discrete-domain process
it is \cite{Riec13b}:
\begin{align*}
\gamma(L) & = \left\langle \overline{\MS}_t \MS_{t+L} \right\rangle_t 
  ~,
\end{align*}
where $L \in \mathbb{Z}$ and the bar denotes the complex conjugate. The
autocorrelation function is symmetric about $L = 0$, so we can focus on $L \geq 0$. 
For $L = 0$, we simply have: 
\begin{align*}
\left\langle  \overline{\MS}_t \MS_{t}  \right\rangle_t 
  & = \sum_{\ms \in \Abet} |\ms|^2 \Pr(\MS_t = \ms) \\
  & = \sum_{\ms \in \Abet} |\ms|^2 \bra{\pi} T^{(\ms)} \ket{\one}
  ~.
\end{align*}
For $L > 0$, we have:
\begin{align*}
\gamma(L) & = \left\langle  \overline{\MS}_t \MS_{t+L}  \right\rangle_t \\
& = \sum_{\ms \in \Abet} \sum_{\ms' \in \Abet} \overline{\ms} \ms'  \, \Pr(\MS_t = \ms, \MS_{t+L} = \ms') \\
& = \sum_{\ms \in \Abet} \sum_{\ms' \in \Abet} \overline{\ms} \ms'  \, \Pr(\ms \underbrace{* \dots *}_{L-1 \text{ } * \text{s}} \ms') \\
&= \sum_{\ms \in \Abet} \sum_{\ms' \in \Abet} \overline{\ms} \ms' \sum_{w \in \Abet^{L-1}} \Pr(\ms w \ms')
  ~.
\end{align*}
Each `$*$' above is a wildcard symbol denoting indifference to the particular
symbol observed in its place. That is, the $*$s denote marginalizing over the
intervening random variables. We develop the consequence of this, explicitly
calculating~\footnote{
Averaging over $t$ invokes unconditioned word probabilities 
that must be calculated using the stationary probability $\pi$ over the recurrent states.
Effectively this ignores any transient nonstationarity that may exist in a process,
since only the recurrent part of the HMM presentation plays a role in the
autocorrelation function.  One practical lesson is that if $T$ has transient states, 
they might as well be trimmed prior to such a calculation.
}
and finding:
\begin{align*}
\gamma(L) &= \sum_{\ms \in \Abet} \sum_{\ms' \in \Abet} \overline{\ms} \ms' \sum_{w \in \Abet^{L-1}} \bra{\pi} T^{(\ms)} T^{(w)} T^{(\ms')} \ket{\one} \\
&= \sum_{\ms \in \Abet} \sum_{\ms' \in \Abet} \overline{\ms} \ms' \bra{\pi} T^{(\ms)} \Bigl(\sum_{w \in \Abet^{L-1}} T^{(w)} \Bigr) T^{(\ms')} \ket{\one} \\
&= \sum_{\ms \in \Abet} \sum_{\ms' \in \Abet} \overline{\ms} \ms' \bra{\pi} T^{(\ms)} \Bigl(\prod_{i=1}^{L-1} \underbrace{\sum_{\ms_i \in \Abet} T^{(\ms_i)} }_{ = T} \Bigr) T^{(\ms')} \ket{\one} \\
&= \sum_{\ms \in \Abet} \sum_{\ms' \in \Abet} \overline{\ms} \ms' \bra{\pi} T^{(\ms)} T^{L-1} T^{(\ms')} \ket{\one} \\
&=   \bra{\pi} \Big( \sum_{\ms \in \Abet} \overline{\ms} T^{(\ms)} \Big) T^{L-1} \Big( \sum_{\ms' \in \Abet}  \ms' T^{(\ms')} \Big) \ket{\one} 
  ~.
\end{align*}

The result is the autocorrelation in cascading form $\braket{\cdot | T^t |
\cdot}$, which can be made particularly transparent by subsuming
time-independent factors on the left and right into the bras and kets. Let's
introduce the new row vector:
\begin{align*}
\corrbra & = \bra{\pi} \Big( \sum_{\ms \in \Abet} \overline{\ms} T^{(\ms)} \Big)
\end{align*}
and column vector:
\begin{align*}
\corrket & = \Big( \sum_{\ms \in \Abet}  \ms T^{(\ms)} \Big) \ket{\one} 
  ~.
\end{align*}
Then, the autocorrelation function for nonzero integer $\tau$ is simply:
\begin{align}
\gamma(L) & = \corrbra T^{|L|-1} \corrket~.
\end{align}
Clearly, the autocorrelation function is a direct, albeit filtered, signature
of iterates of the transition dynamic of any process presentation.

This result can easily be translated to the continuous-time setting. If the
process is represented as a function of a Markov chain and we make the
translation that:
\begin{align*}
\corrbra & = \bra{\pi} \Big( \sum_{\ms \in \Abet} \overline{\ms} \Gamma_{\ms} \Big)
\; \text{ and } \;
\corrket = \Big( \sum_{\ms \in \Abet}  \ms \Gamma_{\ms} \Big) \ket{\one} ~,
\end{align*}
then the autocorrelation function for any $\tau \in \mathbb{R}$ is simply:
\begin{align}
\gamma(\tau) & = \corrbra e^{|\tau| G} \corrket
  ~,
\end{align}
where $G$ is determined from $T$ following \S \ref{sec:ContTimeHMMs}. Again,
the autocorrelation function is a direct fingerprint of the transition dynamic
over the hidden states.

The power spectrum is a modulated accumulation of the autocorrelation function.
With some algebra, one can show that it is:
\begin{align*}
\Psp 
& = \lim_{N \to \infty} \frac{1}{N} \sum_{L=-N}^{N} 
    \bigl( N - |L| \bigr) \, \gamma(L) \, e^{-i \omega L} ~.
\end{align*}
Reference \cite{Riec13b} showed that for discrete-domain processes the
continuous part of the power spectrum is simply: 
\begin{align}
P_c(\omega) & =  \bigl\langle \left| x \right|^2 \bigr\rangle + 
    2 \, \text{Re} \corrbra \left( e^{i \omega} I - \matHMM \right)^{-1} \corrket 
    ~,
\end{align}
where Re$(\cdot)$ denotes the real part of its argument and $I$ is the identity
matrix. Similarly, for continuous-domain processes one has:
\begin{align}
P_c(\omega) & = 
    2 \, \text{Re} \corrbra \left( i \omega I - G \right)^{-1} \corrket ~.
\end{align}

Although useful, these signatures of pairwise correlation are only first-order
complexity measures. Common measures of complexity that include higher orders
of correlation can also be written in the simple cascading form, but require a
more careful choice of representation.

\subsection{Predictability from a presentation MSP}

For example, any HMM presentation allows us to calculate using
Eq.~(\ref{eq:WordDistHMM}) a process's block entropy:
\begin{align*}
\H(L) = \H[\MS_0 \MS_1 \ldots \MS_{L-1}] ~,
\end{align*}
but at a computational cost $\mathcal{O} \left(|\SSet|^3 L |\Abet|^L \right)$
exponential in $L$, due to the exponentially growing number of words in
$\mathcal{L} \cap \Abet^L$. Consequently, using a general HMM one can neither
directly nor efficiently calculate many key complexity measures, including a
process's entropy rate and excess entropy.

These limitations motivate using more specialized HMM classes. To take one
example, it has been known for some time that a process' entropy rate $\hmu$
can be calculated directly from any of its unifilar presentations~\cite{Ash65}.
Another is that we can calculate the excess entropy directly from a process's
uHMM forward and reverse states \cite{Crut08a,Crut08b}: $\EE =
\I[\Past;\Future] = \I[\FutureCausalState;\PastCausalState]$.

However, efficient computation of \emph{myopic} entropy rates $\hmu(L)$
remained elusive for some time, and we only recently found their closed-form
expression \cite{Crut13a}. The myopic entropy rates are important because they
represent the apparent entropy rate of a process if it is modeled as a finite
Markov order-($L-1$) process---a very common approximation. Crucially, the
difference $\hmu(L) - \hmu$ from the process' true entropy rate is the
\emph{surplus entropy rate} incurred by using an order-$L-1$ Markov
approximation. Similarly, these surplus entropy rates lead directly to not only
an apparent loss of predictability, but errors in inferred physical properties.
These include overestimates of dissipation associated with the surplus entropy
rate assigned to a physical thermodynamic system~\cite{Boyd16d}.

Unifilarity, it turns out, is not enough to calculate a process' $\hmu(L)$
directly. Rather, the \syncMSP\ of any process presentation is what is
required. Let's now develop the closed-form expression for the myopic entropy
rates, following Ref.~\cite{Elli13a}.

The length-$L$ myopic entropy rate is the expected uncertainty in the
$L^\text{th}$ random variable $X_{L-1}$, given the preceding $L-1$ random
variables $X_{0 : L-1}$:
\begin{align}
\hmu(L) & \equiv \H(L) - \H(L-1) \nonumber \\    
  & = \H \left[ X_{0 : L} \middle| \mxst_0 = \pi \right]
  	- \H \left[  X_{0 : L-1} \middle| \mxst_0 = \pi \right]
	  \nonumber \\  
  & = \H \left[ X_{L-1} , \, X_{0 : L-1} \middle| \mxst_0 = \pi \right]
  	- \H \left[  X_{0 : L-1} \middle| \mxst_0 = \pi \right]
  	\nonumber \\ 
  & = \H \left[ X_{L-1} \middle| X_{0 : L-1}, \mxst_0 = \pi \right]
  ~, 
\label{eq: block entropy rate as conditional entropy}
\end{align}
where, in the second line, we explicitly give the condition $\mxst_0 = \pi$
specifying our ignorance of the initial state. That is, without making any
observations we can only assume that the initial distribution $\mxst_0$ over
$\mathcal{M}$'s states is the expected asymptotic distribution $\pi$. For a
mixing ergodic process, for example, even if another distribution $\mxst_{-N} =
\alpha$ was known in distant past, we still have $\bra{\mxst_0} =
\bra{\mxst_{-N}} T^N \to \bra{\pi}$, as $N \to \infty$.

Assuming an initial probability distribution over $\mathcal{M}$'s states, a
given observation sequence induces a particular sequence of updated state
distributions. That is, the \syncMSP($\mathcal{M}$)\ is unifilar regardless of
whether $\mathcal{M}$ is unifilar or not. Or, in other words, given the
\syncMSP's\ unique start state---$\MxSt_0 = \pi$---and a particular realization
$X_{0 : L-1} = w^{L-1}$ of the last $L-1$ random variables, we end up at the
particular mixed state $\MxSt_{L-1} = \mxstw{w^{L-1}} \in \MxSSet$. Moreover,
the entropy of the next observation is uniquely determined by $\mathcal{M}$'s
state distribution, suggesting that Eq.  (\ref{eq: block entropy rate as
conditional entropy}) becomes:
\begin{align*} 
\H \left[ X_{L-1} | \, X_{0 : L-1}, \mxst_0 = \pi \right]  
  & = \H \left[ X_{L-1} | \MxSt_{L-1}, \MxSt_0 = \pi \right]
  ~,
\end{align*}
as proven elsewhere \cite{Elli13a}. Intuitively, conditioning on all of the
past observation random variables is equivalent to conditioning on the random
variable for the state distribution induced by particular observation
sequences.

We can now recast Eq. \eqref{eq: block entropy rate as conditional entropy} in
terms of the \syncMSP, finding:
\begin{align*}
\hmu(L) &= \H  \left[ X_{L-1} \middle| \, \MxSt_{L-1}, \MxSt_0 = \pi \right]  \\
  & = \sum_{\mxst \in \MxSSet} \!
  \Pr( \MxSt_{L-1} = \mxst | \MxSt_0 = \pi ) 
  \H \left[ X_{L-1} \middle| \MxSt_{L-1} = \mxst \right] \\
  & = \sum_{\mxst \in \MxSSet} \!
  \StartMS W^{L-1} \ket{\delta_{\mxst}} \\
  & \quad\quad \times
  - \sum_{\ms \in \Abet} \bra{\delta_\mxst} W^{(\ms)} \ket{\one}
  \log_2 \bra{\delta_\mxst} W^{(\ms)} \ket{\one} \\
  & = \StartMS W^{L-1}  \HWA
  ~.
\end{align*}
Here:
\begin{align*}
\HWA \equiv - \!\! \sum_{\mxst \in \MxSSet} \!
	\ket{\delta_{\mxst}}
    \sum_{\ms \in \Abet} \bra{\delta_\mxst} W^{(\ms)} \ket{\one}
	\log_2 \! \bra{\delta_\mxst} W^{(\ms)} \ket{\one} 
\end{align*}
is simply the column vector whose $i^\text{th}$ entry is the entropy of
transitioning from the $i^\text{th}$ state of \syncMSP. Critically, $\HWA$ is
independent of $L$.

Notice that taking the logarithm of the sum of the entries of the row vector
$\bra{\delta_\mxst} W^{(\ms)}$ via $\bra{\delta_\mxst} W^{(\ms)} \ket{\one}$ is
only permissible since \syncMSP's\ unifilarity guarantees that $W^{(\ms)}$ has
at most one nonzero entry per row. (We also use the familiar convention that $0
\log_2 0 = 0$ \cite{Cove06a}.)

The result is a particularly compact and efficient expression for the
length-$L$ myopic entropy rates:
\begin{align}
h_\mu(L) & = \StartMS W^{L-1} \HWA
  ~.
\label{eq:MyopicEntropyRateMSP}
\end{align}
Thus, all that is required is computing powers of the MSP transition dynamic.
The computational cost $\mathcal{O} (L |\MxSSet|^3)$ is now only linear in
$L$.  Moreover, $W$ is very sparse, especially so with a small alphabet
$\Abet$. And, this means that the computational cost can be reduced even
further via numerical optimization.

With $\hmu(L)$ in hand, the hierarchy of complexity measures that derive from
it immediately follow, including the entropy rate $\hmu$, the excess entropy
$\EE$, and the transient information $\TI$ \cite{Crut01a}. Specifically, we
have:
\begin{align*}
\hmu &= \lim_{L \to \infty} \hmu(L) ~, \\
\EE &= \sum_{L=1}^\infty   \left[ h_\mu(L) - h_\mu \right]  ~, ~\text{and} \\
\TI  &= \sum_{L=1}^\infty L \left[ h_\mu(L) - h_\mu \right] ~.
\end{align*}
The sequel, Part II, discusses these in more detail, introducing their
closed-form expressions. To prepare for this, we must first review the
meromorphic functional calculus, which is needed for working with the above
operators.

\subsection{Continuous time?}

We saw that correlation measures are easily extended to the continuous-time
domain via continuous-time HMMs. Information measures, though, are awkward in
continuous time, although progress has been made recently towards understanding
their structure~\cite{Marz14e,Marz14c}.

\subsection{Synchronization from generator MSP}

If a process' state-space is known, then the \syncMSP\ of the generating model
allows one to track the observation-induced distributions over its states. This
naturally leads to closed-form solutions to informational questions about how
an observer comes to know, or how it \emph{synchronizes} to, the system's states.

To monitor how an observer's knowledge of a process' internal state changes with
increasing measurements we use the \emph{myopic state uncertainty}
$\mathcal{H}(L) = \H[\St_0 | \MS_{-L:0} ]$ \cite{Crut01a}. Expressing it in
terms of the \syncMSP, one finds \cite{Elli13a}:
\begin{align*}
\mathcal{H}(L) & = - \sum_{w \in \Abet^L} \Prob(w)
  \sum_{\st \in \SSet} \Prob(\st|w) \log_2 \Prob(\st|w) \\
  & =
  \sum_{\mxst \in \MxSSet}
  \Prob (\MxSt_L = \mxst | \MxSt_0 = \pi ) \H[\mxst] ~.
\end{align*}
Here, $\H[\mxst]$ is the presentation-state uncertainty specified by the mixed
state $\eta$:
\begin{align}
\H[\mxst] \equiv - \sum_{\st \in \SSet}
  \braket{\eta|\delta_\st} \log_2 \braket{\eta|\delta_\st}
  ~, 
\end{align}
where $\ket{\delta_\st}$ is the length-$|\SSet|$ column vector of all zeros
except for a $1$ at the appropriate index of the presentation-state $\st$.

Continuing, we re-express $\mathcal{H}(L)$ in terms of powers of the \syncMSP\
transition dynamic: 
\begin{align}
\mathcal{H}(L) 
  & = \sum_{\mxst \in \MxSSet}  \StartMS W^{L} \ket{\delta_{\mxst}}
  \H \left[ \mxst \right]
  \nonumber \\ 
  & = \StartMS W^{L} \Hmxst
  ~.
\label{eq:StateUncertaintyMSP}
\end{align}
Here, we defined:
\begin{align*}
\Hmxst \equiv  \sum_{\mxst \in \MxSSet}  | \delta_{\mxst} \rangle 
\, \H \! \left[ \mxst \right] , 
\end{align*}
which is the $L$-independent length-$|\MxSSet|$ column vector whose entries are the appropriately indexed entropies of each mixed state. 

The forms of Eqs.~\eqref{eq:MyopicEntropyRateMSP} and
\eqref{eq:StateUncertaintyMSP} demonstrate that $\hmu(L+1)$ and
$\mathcal{H}(L)$ differ only in the type of information being extracted after
being evolved by the operator: observable entropy $\Hmxst$ or state entropy $\H
\left[ \mxst \right]$, as implicated by their respective kets.  Each of these
entropies decreases as the distributions induced by longer observation
sequences converge to their asymptotic form. If synchronization is achieved,
the latter become delta functions on a single state and the associated
entropies vanish.

Paralleling $\hmu(L)$, there is a complementary hierarchy of complexity
measures that are built from functionals of $\mathcal{H}(L)$. These include
the \emph{asymptotic state uncertainty} $\mathcal{H}$ and \emph{excess 
synchronization information} $\SI'$, to mention only two:  
\begin{align*}
\mathcal{H} &= \lim_{L \to \infty} \mathcal{H}(L) ~\text{and} \\
	\SI' &= \sum_{L=0}^\infty   \left[ \mathcal{H}(L) - \mathcal{H} \right]
	~.
\end{align*}
Compared to the $\hmu(L)$ family of measures, $\mathcal{H}$ and $\SI'$ mirror
the roles of $\hmu$ and $\EE$, respectively.

The \emph{model state-complexity}:
\begin{align*}
C(\mathcal{M}) & = \mathcal{H}(0) \\
  & = \langle \delta_\pi \Hmxst
\end{align*}
also has an analog in the $\hmu(L)$ hierarchy---the process' \emph{alphabet
complexity}:
\begin{align*}
\H[\MS_0] & = \hmu(1) \\
  & =  \langle \delta_\pi \HWA
  ~.
\end{align*}

\subsection{Optimal prediction from \eM\ MSP}

We just reviewed the linear underpinnings of synchronizing to \emph{any} model
of a process. However, the myopic state uncertainty of the $\eM$ has a
distinguished role in determining the synchronization cost for optimally
predicting a process, regardless of the presentation that generated it. Using
the \eM's \syncMSP, the \eM\ myopic state uncertainty can be written in direct
parallel to the myopic state uncertainty of any model:
\begin{align*}
\mathcal{H}^{\forward}(L) & = - \sum_{w \in \Abet^L} \Prob(w)
  \sum_{\cs \in \SSet^{\forward}} \Prob(\cs | w) \log_2 \Prob(\cs | w) \\
  & =
  \sum_{\mxst \in \MxSSet^{\forward}}
  \Prob (\MxSt_L = \mxst | \MxSt_0 = \pi ) \H[\mxst] \\
  &= 
   \StartMS \mathcal{W}^{L} \Hmxst
  ~.
\end{align*}
The script $\mathcal{W}$ emphasizes that we are now specifically working with
the state-to-state transition dynamic of the \eM's MSP.

Paralleling $\mathcal{H}(L)$, an obvious hierarchy of complexity measures is
built from functionals of $\mathcal{H}^{\forward}(L)$. For example, the \eM's
state-complexity is the \emph{statistical complexity} $\Cmu =
\mathcal{H}^{\forward}(0)$. The information that must be obtained to
synchronize to the causal state and thus optimally predict---the \emph{causal
synchronization information}---is given in terms of the \eM's \syncMSP\ by $\SI
= \sum_{L=0}^\infty \mathcal{H}^{\forward}(L)$.

An important difference when using \eM\ presentations is that they have zero
asymptotic state uncertainty:
\begin{align*}
\mathcal{H}^{\forward} = 0 ~.
\end{align*}
Therefore, $\SI = \SI'(\eM)$. Moreover, we conjecture that $\SI =
\min_{\mathcal{M}} \sum_{L=0}^\infty \mathcal{H}(L)$ for any presentation
$\mathcal{M}$ that generates the process, even if $\Cmu \geq C_g$.

\subsection{Beyond the MSP}

Many of the complexity measures use a mixed-state presentation as the
appropriate linear dynamic, with particular focus on the \syncMSP. However, we
want to emphasize that this is more a reflection of questions that have become
common. It does not indicate the general answer that one expects in the broader
approach to finding the hidden linear dynamic. Here, we give a brief overview
for how other linear dynamics can appear for different types of complexity
questions. These have been uncovered recently and will be reported on in more
detail in sequels.

First, we found the reverse-time \emph{mixed-functional presentation} (MFP) of
any forward-time generator. The MFP tracks the reverse-time dynamic over linear
functionals $\ket{\mxst}$ of state distributions induced by reverse-time
observations:
\begin{align*}
\ket{\mxst} \in \MxFSet = \left\{
  \frac{T^{(w)} \ket{\one}}{\bra{\pi} T^{(w)} \ket{\one} }
  \right\}_{w} 
  ~.
\end{align*}
The MFP allows direct calculation of the convergence of the \emph{preparation
uncertainty} $\PU(L) \equiv \H(\St_0 | X_{0:L})$ via powers of the linear MFP
transition dynamic. The preparation uncertainty in turn gives a new perspective
on the transient information since:
\begin{align*}
\TI = \sum_{L=0}^\infty \bigl( \H(\St_0^+ | X_{0:L}) - \PC \bigr)
\end{align*}
can be interpreted as the predictive advantage of hindsight. Related, the
\emph{myopic process crypticity} $\PC(L) = \PU^+(L) - \mathcal{H}^+(L)$ had
been previously introduced~\cite{Maho11a}. Since $\lim_{L \to \infty}
\mathcal{H}^+(L) = \mathcal{H}^+ = 0$, the asymptotic crypticity is $\PC =
\PU^+ + \mathcal{H}^+ = \PU^+$. And, this reveals a refined partitioning
underlying the sum:
\begin{align*}
\sum_{L=0}^\infty \bigl( \PC - \PC(L) \bigr) = \SI - \TI \geq 0
  ~.
\end{align*}

Crypticity $\PC = \H(\St_0^+ | X_{0: \infty}) $ itself is positive only if the process' \emph{cryptic order}:
\begin{align*}
k = \min \bigl\{ \ell \in \{ 0, 1, \dots \} : \H(\St_0^+ | X_{-\ell: \infty}) = 0 \bigr\} ~,
\end{align*}
is positive. The cryptic order is always less than or equal to its better known
cousin, the \emph{Markov order} $\MOrder$:
\begin{align*}
\MOrder = \min \bigl\{ \ell \in \{ 0, 1, \dots \} : \H(\St_0^+ | X_{-\ell:0}) = 0 \bigr\}
  ~,
\end{align*}
since conditioning can never increase entropy. In the case of cryptic order, we condition on future observations $X_{0:\infty}$.

The forward-time \emph{cryptic operator presentation} gives the forward-time
observation-induced dynamic over the operators:
\begin{align*}
\mathcal{O} \in \left\{
  \frac{ \ket{s^-} \bra{\mxst^w} }{\braket{\mxst^w | s^- } } :
  s^- \in \St^-, \bra{\mxst^w} \in \MxSSet , \braket{\mxst^w | s^-} > 0 
  \right\}
  ~.
\end{align*}
Since the reverse causal state $\St_0^-$ at time 0 is a linear combination of
forward causal states~\cite{Maho09a, Maho09b}, this presentation allows new
calculations of the convergence to crypticity that implicate $\Pr(\St_0^+ |
X_{-L: \infty})$.

In fact, the cryptic operator presentation is a special case of the more general
myopic bidirectional dynamic over operators :
\begin{align*}
\mathcal{O} \in \left\{
  \frac{ \ket{\mxst^{w'}} \bra{\mxst^w} }{\braket{\mxst^w | \mxst^{w'} } } :
  \bra{\mxst^w} \in \MxSSet ,
  \ket{\mxst^{w'}} \in \MxFSet ,  \braket{\mxst^w | \mxst^{w'} } > 0
  \right\}
\end{align*}
induced by new observations of either the future or the past. This is key to
understanding the interplay between forgetfulness and shortsightedness:
$\Pr(\St_0 | X_{-M:0}, X_{0:N})$.

The list of these extensions continues. Detailed bounds on entropy-rate
convergence are obtained from the transition dynamic of the so-called
\emph{possibility machine}, beyond the asymptotic result obtained in
Ref.~\cite{Trav10b}. And, the importance of post-synchronized monitoring, as
quantified by the information lost due to negligence over a duration $\ell$:
\begin{align*}
\bmu(\ell) = \I(X_{0:\ell} ; X_{\ell: \infty} | X_{-\infty: 0})
  ~,
\end{align*}
can be determined using yet another type of modified MSP.

These examples all find an exact solution via a theory parallel to that
outlined in the following, but applied to the linear dynamic appropriate for
the corresponding complexity question. Furthermore, they highlight the
opportunity, enabled by the full meromorphic functional calculus
\cite{Riec16a}, to ask and answer more nuanced and, thus, more probing
questions about structure, predictability, and prediction.

\subsection{The end?}

It would seem that we achieved our goal. We identified the appropriate
transition dynamic for common complexity questions and, by some standard, gave
formulae for their exact solution. In point of fact, the effort so far has all
been in preparation. Although we set the framework up appropriately for linear
analysis, closed-form expressions for the complexity measures still await the
mathematical developments of the following sections. At the same time, at the
level of qualitative understanding and scientific interpretation, so far we
failed to answer the simple question:
\begin{itemize}
\setlength{\itemsep}{-4pt}
\setlength{\topsep}{-6pt}
\setlength{\parsep}{-6pt}
\setlength\itemindent{-15pt}
\item What range of possible behaviors do these complexity measures exhibit?
\end{itemize}
and the natural follow-up question:
\begin{itemize}
\setlength{\itemsep}{-4pt}
\setlength{\topsep}{-6pt}
\setlength{\parsep}{-6pt}
\setlength\itemindent{-15pt}
\item What mechanisms produce qualitatively different informational signatures?
\end{itemize}

The following section reviews the recently developed functional calculus that
allows us to actually decompose arbitrary functions of the nondiagonalizable
hidden dynamic to give conclusive answers to these fundamental questions
\cite{Riec16a}. We then analyze the range of possible behaviors and identify
the internal mechanisms that give rise to qualitatively different contributions
to complexity.

The investment in this and the succeeding sections allow Part II to express new
closed-form solutions for many complexity measures beyond what those achieved
to date. In addition to obvious calculational advantages, this also gives new
insights into possible behaviors of the complexity measures and, moreover,
their unexpected similarities with each other. In many ways, the results shed
new light on what we were (implicitly) probing with already-familiar complexity
measures. Constructively, this suggests extending complexity magnitudes to
complexity functions that succinctly capture the organization to all orders of
correlation. Just as our intuition for pairwise correlation grows out of power
spectra, so too these extensions unveil the workings of both a process'
predictability and the burden of prediction for an observer.

\section{Spectral Theory beyond the Spectral Theorem}
\label{sec:SpectralTheory}

Here, we briefly review the spectral decomposition theory from
Ref.~\cite{Riec16a} needed for working with linear operators. As will become
clear, it goes significantly beyond the spectral theorem for normal operators. 

\subsection{Spectral primer}
\label{sec:SpectralPrimer}

We restrict our attention to operators that have at most a countably infinite
spectrum. Such operators share many features with finite-dimensional square
matrices. And so, we review several elementary but essential facts that are
used extensively in the following.

Recall that if $\opGen$ is a finite-dimensional square matrix, then $\opGen$'s
spectrum is simply its set of eigenvalues:
\begin{align*}
\Lambda_\opGen = \bigl\{ \lambda \in \mathbb{C}: \text{det}(\lambda I - \opGen) = 0 \bigr\}
  ~,
\end{align*}
where det$(\cdot)$ is the determinant of its argument.

For reference later, recall that the \emph{algebraic multiplicity} $a_\lambda$
of eigenvalue $\lambda$ is the power of the term $(z-\lambda)$ in the
characteristic polynomial det$(zI - \opGen)$. In contrast, the \emph{geometric
multiplicity} $g_\lambda$ is the dimension of the kernel of the transformation
$\opGen - \lambda I$ or the number of linearly independent eigenvectors for the
eigenvalue. The algebraic and geometric multiplicities are all equal when the
matrix is diagonalizable.

Since there can be multiple subspaces associated with a single eigenvalue,
corresponding to different Jordan blocks in the Jordan canonical form, it is
structurally important to introduce the index of the eigenvalue to describe the
size of its largest-dimension associated subspace.

\begin{Def} 
The \emph{index} $\nu_\lambda$ of eigenvalue $\lambda$ is the size of the largest Jordan block associated with $\lambda$.  
\end{Def} 

The index gives information beyond what the algebraic and geometric
multiplicities themselves reveal. Nevertheless, for $\lambda \in \Lambda_A$, it
is always true that $\nu_\lambda - 1 \leq a_\lambda - g_\lambda \leq
a_\lambda - 1$. In the diagonalizable case, $a_\lambda = g_\lambda$ and
$\nu_\lambda = 1$ for all $\lambda \in \Lambda_A$.

The \emph{resolvent}:
\begin{align*}
\R(\z; \opGen) \equiv (z I - \opGen)^{-1} ~,
\end{align*}
defined with the help of the continuous complex variable $\z \in \mathbb{C}$,
captures all of the spectral information about $\opGen$ through the poles of
the resolvent's matrix elements. In fact, the resolvent contains more than just
the spectrum: the order of each pole gives the index of the corresponding
eigenvalue.

Each eigenvalue $\lambda$ of $A$ has an associated \emph{projection operator}
$A_\lambda$, which is the residue of the resolvent as $z \to \lambda$:
\begin{align}
A_{\lambda} \equiv
\frac{1}{2 \pi i} \oint_{C_{\lambda}} \R(\z; \opGen)  d \z
\label{eq:ProjOpsViaRes}
~.
\end{align}
The residue of the matrix can be calculated elementwise.

The projection operators are orthonormal:
\begin{align}
\label{eq:ProjOpsAreOrthonormal}
\opGen_\lambda \opGen_\zeta = \delta_{\lambda, \zeta} \opGen_\lambda
  ~,
\end{align} 
and sum to the identity: 
\begin{align}
I  & = \sum_{\lambda \in \Lambda_\opGen} \opGen_{\lambda}
  ~. 
\label{eq:ProjOpsSumToIdentity}
\end{align} 

For cases where $\nu_\lambda = 1$, we found that the projection operator
associated with $\lambda$ can be calculated as \cite{Riec16a}:
\begin{align} 
\opGen_\lambda & =
  \prod_{\zeta \in \Lambda_\opGen \atop \zeta \neq \lambda}
 	\left( \frac{\opGen - \zeta I }{\lambda - \zeta} \right)^{\nu_\zeta} 
  ~. 
\label{eq: proj operators for index-one eigs}
\end{align}
Not all projection operators of a nondiagonalizable operator can be found
directly from Eq.~\eqref{eq: proj operators for index-one eigs}, since some
have index larger than one. However, if there is only one eigenvalue that has
index larger than one---the \emph{almost diagonalizable} case treated in Part
II---then
Eq.~\eqref{eq: proj operators for index-one eigs}, together with the fact that
the projection operators must sum to the identity, \emph{does} give a full
solution to the set of projection operators. Next, we consider the general
case, with no restriction on $\nu_\lambda$.

\subsection{Eigenprojectors: Left, right, generalized}
\label{sec:GenEigvects}

In general, as we now discuss, an operator's eigenprojectors can be obtained
from all left and right eigenvectors and generalized eigenvectors associated
with the eigenvalue. Given the $n$-tuple of possibly-degenerate eigenvalues
$(\Lambda_A) = (\lambda_1, \, \lambda_2, \, \dots \, , \, \lambda_n )$, there
is a corresponding $n$-tuple of $m_k$-tuples of linearly-independent
\emph{generalized right-eigenvectors}:
\begin{align*}
\left( ( \ket{\lambda_1^{(m)}} )_{m=1}^{m_1} , \, ( \ket{\lambda_2^{(m)}} )_{m=1}^{m_2}, \, \dots \, , \, ( \ket{\lambda_n^{(m)}} )_{m=1}^{m_n} \right)
  ~,
\end{align*}
where:
\begin{align*}
( \ket{\lambda_k^{(m)}} )_{m=1}^{m_k} \equiv \left( \ket{\lambda_k^{(1)}} , \, \ket{\lambda_k^{(2)}} , \, \dots \, , \,\ket{\lambda_k^{(m_k)}} \right) 
\end{align*}
and a corresponding $n$-tuple of $m_k$-tuples of linearly-independent
\emph{generalized left-eigenvectors}:
\begin{align*}
\left( ( \bra{\lambda_1^{(m)}} )_{m=1}^{m_1} , \, ( \bra{\lambda_2^{(m)}} )_{m=1}^{m_2}, \, \dots \, , \, ( \bra{\lambda_n^{(m)}} )_{m=1}^{m_n} \right)
  ~,
\end{align*}
where:
\begin{align*}
( \bra{\lambda_k^{(m)}} )_{m=1}^{m_k} \equiv \left( \bra{\lambda_k^{(1)}} , \, \bra{\lambda_k^{(2)}} , \, \dots \, , \,\bra{\lambda_k^{(m_k)}} \right) 
\end{align*}
such that:
\begin{align}
(A - \lambda_k I ) \ket{\lambda_k^{(m+1)}} =  \ket{\lambda_k^{(m)}} 
\label{eq:RightGenRecursion}
\end{align}
and:
\begin{align}
\bra{\lambda_k^{(m+1)}} (A - \lambda_k I ) = \bra{\lambda_k^{(m)}}  
  ~,
\label{eq:LeftGenRecursion}
\end{align}
for $0 \leq m \leq m_k - 1$, where $\ket{\lambda_j^{(0)}} = \vec{0}$ and
$\bra{\lambda_j^{(0)}} = \vec{0}$. Specifically, $\ket{\lambda_k^{(1)}}$ and
$\bra{\lambda_k^{(1)}}$ are conventional right and left eigenvectors,
respectively.

Recall that eigenvalue $\lambda \in \Lambda_A$ corresponds to $g_\lambda$
different Jordan blocks, where $g_\lambda$ is $\lambda$'s geometric
multiplicity. In fact:
\begin{align*}
n = \sum_{\lambda \in \Lambda_A} g_\lambda ~.
\end{align*}
Moreover, $\lambda$'s index $\nu_\lambda$ is the size of the largest Jordan
block corresponding to $\lambda$:
\begin{align*}
\nu_\lambda = \max \{ \delta_{\lambda, \lambda_k} m_k \}_{k=1}^n
  ~.
\end{align*}

Most directly, the generalized right and left eigenvectors can be found as the
nontrivial solutions to:
\begin{align*}
(A - \lambda_k I )^m \ket{\lambda_k^{(m)}} =  \ket{0} 
\end{align*}
and:
\begin{align*}
\bra{\lambda_k^{(m)}} (A - \lambda_k I )^m = \bra{0}
  ~,
\end{align*}
respectively.
Imposing appropriate normalization, we find that:
\begin{align}
\braket{\lambda_j^{(m)} | \lambda_k^{(n)}} = \delta_{j, k} \delta_{m + n, m_k + 1} ~.
\label{eq:GenEigenvectorOrthogonality}
\end{align}

Crucially, right and left eigenvectors are no longer simply related by complex-conjugate transposition and right eigenvectors are not necessarily orthogonal to each other. Rather, left eigenvectors and generalized eigenvectors form a dual basis to the right eigenvectors and generalized eigenvectors. Somewhat surprisingly, the \emph{most} generalized left eigenvector $\bra{ \lambda_k^{(m_k)}}$ associated with $\lambda_k$ is dual to the \emph{least} generalized right eigenvector $\ket{ \lambda_k^{(1)}}$ associated with $\lambda_k$:
\begin{align*}
\braket{ \lambda_k^{(m_k)} |  \lambda_k^{(1)} } = 1
  ~.
\end{align*}

Explicitly, we find that 
the projection operators for a nondiagonalizable matrix can be written as: 
\begin{align} 
\opGen_{\lambda} = \sum_{k=1}^n \sum_{m = 1}^{m_k} 
  \delta_{\lambda, \lambda_k} \ket{\lambda_k^{(m)}} \bra{\lambda_k^{(m_k + 1 - m)}} ~.
\label{eq:ProjectorsViaGenEigenvectors}
\end{align}

\subsection{Companion operators and resolvent decomposition}

It is useful to introduce the generalized set of \emph{companion operators}:
\begin{align} 
\opGen_{\lambda, m} 
  & = \opGen_\lambda \bigl( \opGen -  \lambda I \bigr)^m 
  ~,
\label{eq:ProjOpExpression4ResidueMatrices}
\end{align} 
for $\lambda \in \Lambda_A$ and $m \in \{ 0, 1, 2, \dots \}$.
These operators satisfy the following semigroup relation:
\begin{align}
\opGen_{\lambda, m} \opGen_{\zeta, n} = \delta_{\lambda, \zeta} \opGen_{\lambda, m+n}
  ~.
\label{eq:GenMatrixOrthogonalityRelation}
\end{align} 
$A_{\lambda, m}$
reduces to the eigenprojector for $m=0$:
\begin{align}
A_{\lambda, 0} = A_\lambda ~,
\end{align}
and it exactly reduces to the zero-matrix for $m \geq \nu_\lambda$:
\begin{align}
\opGen_{\lambda, m} &= \mathbf{0}
  ~. 
\end{align}
Crucially, we can rewrite the resolvent as a weighted sum of the companion
matrices $\{ A_{\lambda, m} \}$, with complex coefficients that have poles at
each eigenvalue $\lambda$ up to the eigenvalue's index $\nu_\lambda$:
\begin{align} 
\R(\z; \opGen) & = 
  \sum_{\lambda \in \Lambda_\opGen} \sum_{m = 0}^{\nu_\lambda - 1}
  \frac{1}{(\z - \lambda)^{m+1}}  A_{\lambda,m} ~.
  \label{eq:PartialFractionsExpansion_of_Resolvent_2}
\end{align}  

Ultimately these results allow us to evaluate arbitrary functions of
nondiagonalizable operators, to which we now turn. (Reference~\cite{Riec16a}
gives more background.)

\subsection{Functions of nondiagonalizable operators}

The \emph{meromorphic functional calculus}~\cite{Riec16a} gives meaning to
arbitrary functions $f(\cdot)$ of any linear operator $A$. Its starting point
is the Cauchy-integral-like formula:
\begin{align}
f(\opGen) & = \sum_{\lambda \in \Lambda_\opGen}
  \frac{1}{2 \pi i} \oint_{C_{\lambda}} f(\z) \R(\z; \opGen)  d \z 
  ~,
\label{eq:PartlyDecomposedCauchyIntegralFormula}
\end{align} 
where $C_\lambda$ denotes a sufficiently small counterclockwise contour 
around $\lambda$ in the complex plane such that no singularity of the integrand besides the possible pole at $z = \lambda$ is enclosed by the contour.

Invoking Eq.~\eqref{eq:PartialFractionsExpansion_of_Resolvent_2} yields the desired formulation:
\begin{align}
f(A)  & = \sum_{\lambda \in \Lambda_\opGen} \sum_{m = 0}^{\nu_\lambda - 1}
      \opGen_{\lambda, m} \, 
      \frac{1}{2 \pi i} \oint_{C_\lambda} \frac{f(z)}{(z - \lambda)^{m+1}} \, dz  ~.
   \label{eq:MFC_desired_formulation}
\end{align}   
Hence, with the eigenprojectors $\{ A_\lambda \}_{\lambda \in \Lambda_A}$ in hand, 
evaluating an arbitrary function of the nondiagonalizable operator $A$
comes down to the evaluation of several residues.

Typically, evaluating Eq.~\eqref{eq:MFC_desired_formulation} requires less work
than one might expect when looking at the equation in its full generality. For
example, whenever $f(z)$ is holomorphic (i.e., well behaved) at $z = \lambda$,
the residue simplifies to:
\begin{align*}
\frac{1}{2 \pi i} \oint_{C_\lambda} \frac{f(z)}{(z - \lambda)^{m+1}} \, dz = \frac{1}{m!} f^{(m)}(\lambda)
  ~,
\end{align*}
where $f^{(m)}(\lambda)$ is the $m^\text{th}$ derivative of $f(z)$ evaluated at
$z = \lambda$. However, if $f(z)$ has a pole or zero at $z=\lambda$, then it
substantially changes the complex contour integration. In the simplest case,
when $A$ is diagonalizable and $f(z)$ is holomorphic at $\Lambda_A$, the
matrix-valued function reduces to the simple form:
\begin{align*}
f(A) = \sum_{\lambda \in \Lambda_A} f(\lambda) \, A_\lambda
  ~.
\end{align*}
Moreover, if $\lambda$ is nondegenerate, then:
\begin{align*}
A_\lambda = \frac{\ket{\lambda} \bra{\lambda}}{ \braket{\lambda | \lambda } }
~,
\end{align*}
although $\bra{\lambda}$ here should be interpreted as the solution to the left
eigenequation $\bra{\lambda} A = \lambda \bra{\lambda}$ and, in general,
$\bra{\lambda} \neq (\ket{\lambda})^\dagger$.

The meromorphic functional calculus agrees with the Taylor-series approach
whenever the series converges and agrees with the holomorphic functional
calculus of Ref.~\cite{Dunford43} whenever $f(z)$ is holomorphic at
$\Lambda_A$.  However, when both these functional calculi fail, the meromorphic
functional calculus extends the domain of $f(A)$ in a way that is key to the
following analysis. We show, for example, that within the meromorphic
functional calculus, the negative-one power of a singular operator is the
Drazin inverse. The Drazin inverse effectively inverts everything that is
invertible. Notably, it appears ubiquitously in the new-found solutions to many
complexity measures.

\subsection{Evaluating residues}

How does one use Eq.\ \eqref{eq:MFC_desired_formulation}? It says that the
spectral decomposition of $f(\opGen)$ reduces to the evaluation of several
residues, where:
\begin{align*}
\text{Res} \bigl(  g(z) , \;  z \to\lambda  \bigr) = \frac{1}{2 \pi i} \oint_{C_\lambda} g(z) \, dz
  ~. 
\end{align*}
So, to make progress with Eq.\ \eqref{eq:MFC_desired_formulation}, we must
evaluate functional-dependent residues of the form $\text{Res} \left( f(z) /
(z - \lambda)^{m+1} , \, z \to \lambda \right)$.  This is basic complex
analysis. Recall that the residue of a complex-valued function $g(z)$ around
its isolated pole $\lambda$ of order $n+1$ can be calculated from:
\begin{align*}
\text{Res} \bigl(  g(z) , \;  z \to\lambda  \bigr) 
  & = \frac{1}{n!} \,  \lim_{z \to \lambda} \, \frac{d^{n}}{{dz}^{n}}
  \left[  (z - \lambda)^{n+1}  g(z)  \right] 
  ~.
\end{align*} 

\subsection{Decomposing $\opGen^L$}

Equation \eqref{eq:MFC_desired_formulation} allows us to explicitly derive the
spectral decomposition of powers of an operator. For $f(\opGen) = \opGen^L \to
f(z) = z^L$, $z=0$ can be either a zero or a pole of $f(z)$, depending on the
value of $L$. In either case, an eigenvalue of $\lambda=0$ will distinguish
itself in the residue calculation of $A^L$ via its unique ability to change the
order of the pole (or zero) at $z=0$.

For example, at this special value of $\lambda$ and for integer $L > 0$,
$\lambda = 0$ induces poles that \emph{cancel} with the zeros of $f(z) = z^L$,
since $z^L$ has a zero at $z=0$ of order $L$. For integer $L < 0$, an
eigenvalue of $\lambda = 0$ \emph{increases} the order of the $z=0$ pole of
$f(z) = z^L$. For all other eigenvalues, the residues will be as expected.

Hence, for any $L \in \mathbb{C}$: 
\begin{align} 
\opGen^L & = \Biggl[
  \sum_{\lambda \in \Lambda_\opGen \atop \lambda \neq 0}
  \sum_{m = 0}^{\nu_\lambda - 1}
  \binom{L}{m} \lambda^{L-m}  
  \opGen_{\lambda, m} \Biggr] \nonumber \\
  & \qquad + \left[ 0 \in \Lambda_\opGen \right] 
  \sum_{m = 0}^{\nu_0 - 1} \delta_{L, m}  \opGen_0  \opGen^m 
  ~,
\label{eq: T^n generally}
\end{align}
where $\binom{L}{m}$ is the generalized binomial coefficient:
\begin{align}
\binom{L}{m} & = \frac{1}{m!} \prod_{n=1}^m (L-n+1)
  ~,
\end{align}
with $\binom{L}{0} = 1$ and where $[ 0 \in \Lambda_\opGen ]$ is the Iverson
bracket. The latter takes value $1$ if $0$ is an eigenvalue of $\opGen$ and
value $0$ if not. Equation \eqref{eq: T^n generally} applies to any linear
operator with only isolated singularities in its resolvent.

If $L$ is a nonnegative integer such that $L \ge \nu_\lambda - 1$ for all 
$\lambda \in \Lambda_\opGen$, then: 
\begin{align}
\opGen^L 
    & =   \sum_{\lambda \in \Lambda_\opGen \atop \lambda \neq 0} 
        \sum_{m=0}^{\nu_\lambda - 1}  \binom{L}{m} \lambda^{L-m}
		\opGen_{\lambda, m} 
  ~,
  \label{eq: W^L spectral decomp for positive integer L}		
\end{align}
where $\binom{L}{m}$ is now reduced to the traditional binomial coefficient
$L! / m! (L-m)!$.

\subsection{Drazin inverse}

\newcommand{\inv}{\text{inv}}

The negative-one power of a linear operator is in general \emph{not} the same
as the inverse $\inv (\cdot)$, since $\inv (\opGen)$ need not exist. However,
the negative-one power of a linear operator is always defined via Eq.\
\eqref{eq: T^n generally}:
\begin{align}
\opGen^{-1}
  & = \sum_{\lambda \in \Lambda_\opGen \setminus \{0\} } 
		\sum_{m=0}^{\nu_\lambda - 1} 
		  (-1)^m \lambda^{- 1 - m}  \opGen_{\lambda, m} 
\label{eq:Neg1Power}
  ~. 	
\end{align}
Notably, when the operator is singular, we find that:
\begin{align*}
\opGen \opGen^{-1} = I - \opGen_0
  ~.
\end{align*}

This is the \emph{Drazin inverse} $A^\mathcal{D}$ of $\opGen$, also known as
the $\{ 1^{\nu_0}, 2, 5\}$-inverse~\cite{BenIs03}. (Note that it is \emph{not}
the same as the Moore--Penrose pseudo-inverse.) Although the Drazin inverse is
usually defined axiomatically to satisfy certain criteria, here it naturally
derived as the negative one power of a singular operator in the meromorphic
functional calculus.

Whenever $\opGen$ \emph{is} invertible, however, $\opGen^{-1} = \inv (\opGen)$.
That said, we should not confuse this coincidence with equivalence. More to the
point, there is no reason other than accidents of historic notation that the
negative-one power should in general be equivalent to the inverse---especially
if an operator is not invertible. To avoid confusing $\opGen^{-1}$ with $\inv
(\opGen)$, we use the notation $\opGen^{\mathcal{D}}$ for the Drazin inverse of
$\opGen$. Still, $\opGen^{\mathcal{D}} = \text{inv}(\opGen)$ whenever $0 \notin
\Lambda_{\opGen}$.

Although Eq.~\eqref{eq:Neg1Power} is a constructive way to build the Drazin
inverse, it suggests more work than is actually necessary. We derived several
simple constructions for it that require only the original operator and the
eigenvalue-$0$ projector. For example, Ref.~\cite{Riec16a} found that, for any
$c \in \mathbb{C} \setminus \{ 0 \}$:
\begin{align}
A^{\mathcal{D}} &= (I - A_0) (A + cA_0)^{-1}
  ~.
\label{eq:Drazin_from_simple_product_w_cA0}
\end{align}

Later, we will also need the decomposition of $(I-W)^{\mathcal{D}}$, as it
enters into many closed-form complexity expressions. Reference~\cite{Riec16a}
showed that:
\begin{align}
(I - T)^{\mathcal{D}} & = \left[ I - (T - T_1) \right]^{-1} - T_1 
\label{eq: Drazin vs I less Q}
\end{align}
for any stochastic matrix $T$.
If $T$ is the state-transition matrix of an ergodic process, then the RHS of
Eq.~\eqref{eq: Drazin vs I less Q} becomes especially simple to evaluate since then $T_1 = \ket{\one} \bra{\pi}$.

Somewhat tangentially, this connects to the \emph{fundamental matrix} $Z = (I -
T + T_1)^{-1}$ used by Kemeny and Snell~\cite{Keme60} in their analysis of
Markovian dynamics. More immediately, Eq.~\eqref{eq: Drazin vs I less Q} plays
a prominent role when deriving excess entropy and synchronization information.
The explicit spectral decomposition is also useful:
\begin{align}
(I - T)^{\mathcal{D}} & = 
 \!\!\! \sum_{\lambda \in \Lambda_T \setminus \{ 1 \} } 
    \sum_{m=0}^{\nu_\lambda - 1} 
    \frac{1}{(1-\lambda)^{m+1}} 
    T_{\lambda, m}
	~.
\label{eq:ResolventAt1}
\end{align}

\section{Projection Operators for Stochastic Dynamics} 
\label{sec:ProjOpsForStochasticMatrices}

The preceding employed the notation that $A$ is a general linear operator. In
the following, we reserve $T$ for the operator of a stochastic transition
dynamic, as in the state-to-state transition dynamic of an HMM: $T = \sum_{x
\in \Abet} T^{(x)}$. If the state space is finite and has a stationary
distribution, then $T$ has a representation that is a nonnegative
row-stochastic---all rows sum to unity---transition matrix.

We are now in a position to summarize several useful properties for the
projection operators of any row-stochastic matrix $\matHMM$. Naturally, if one
uses column-stochastic instead of row-stochastic matrices, all results can be
translated by simply taking the transpose of every line in the derivations.
(Recall that $(ABC)^\top = C^\top B^\top A^\top$.)

The transition matrix's nonnegativity guarantees that for each $\lambda \in
\Lambda_T$ its complex conjugate $\overline{\lambda}$ is also in $\Lambda_T$.
Moreover, the projection operator associated with the complex conjugate of
$\lambda$ is the complex conjugate of $T_\lambda$:
\begin{align*}
\matHMM_{\overline{\lambda}} = \overline{\matHMM_{\lambda}}
  ~.
\end{align*}

If the dynamic induced by $T$ has a stationary distribution over the state
space, then $T$'s spectral radius is unity and all its eigenvalues lie on or
within the unit circle in the complex plane. The maximal eigenvalues have unity
magnitude and $1 \in \Lambda_T$. Moreover, an extension of the
Perron--Frobenius theorem guarantees that eigenvalues on the unit circle have
algebraic multiplicity equal to their geometric multiplicity. And, so,
$\nu_\zeta = 1$ for all $\zeta \in \{ \lambda \in \Lambda_T:  | \lambda | = 1
\}$.

$T$'s index-one eigenvalue $\lambda=1$ is associated with stationarity of the
hidden Markov chain. $T$'s other eigenvalues on the unit circle are roots of
unity and correspond to deterministic periodicities within the process.

\begin{figure}
\centering
\subfloat[]{\label{subfig:CyclicCluster}
\includegraphics[width=.5\columnwidth]{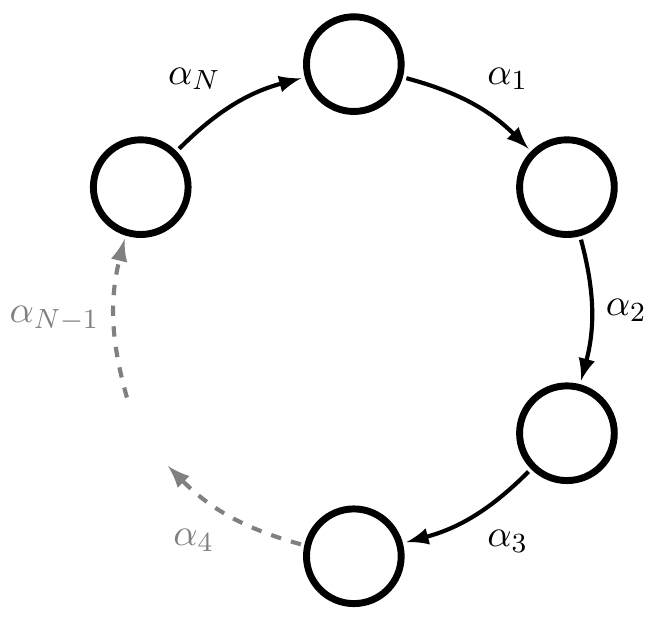} }
\subfloat[]{\label{subfig:DoublyCyclicCluster}
\includegraphics[width=.5\columnwidth]{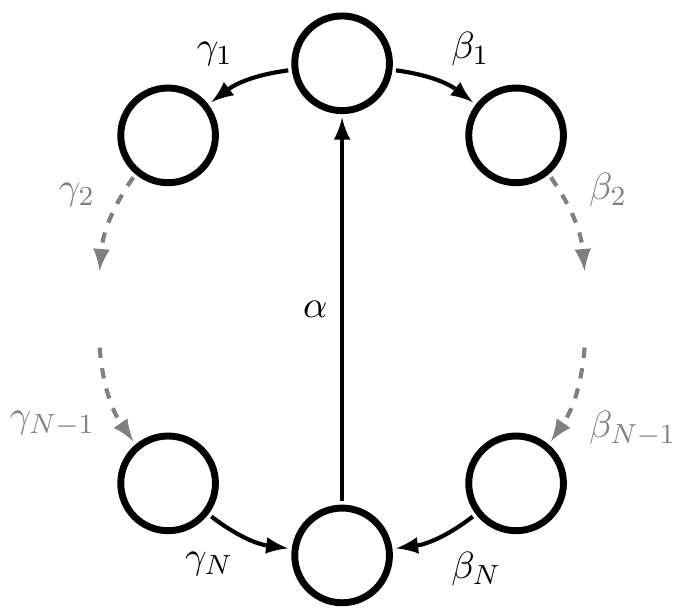} }
\caption{(a) Weighted directed graph (digraph) of the feedback matrix $A$ of a
	cyclic cluster structure that contributes eigenvalues $\Lambda_A = \Bigl\{
	\bigl( \prod_{i=1}^N \alpha_i \bigr)^{1/N} e^{i n 2 \pi / N}
	\Bigr\}_{n=0}^{N-1}$ with algebraic multiplicities $a_\lambda = 1$ for all
	$\lambda \in \Lambda_A$.
	(b) Weighted digraph of the feedback matrix $A$ of a doubly cyclic cluster
	structure that contributes eigenvalues $\Lambda_A = \bigl\{ 0 \bigr\} \cup
	\biggl\{  \Bigl( \alpha \bigl[ \bigl( \prod_{i=1}^N \beta_i \bigr) + \bigl(
	\prod_{i=1}^N \gamma_i \bigr) \bigr] \Bigr)^{ \frac{1}{N+1} } e^{ \frac{i n
	2 \pi }{N+1} } \Bigr\}_{n=0}^{N}$ with algebraic multiplicities $a_0 = N-1$
	and $a_\lambda = 1$ for $\lambda \neq 0$.
	(This eigenvalue ``rule'' depends on having the same number of
	$\beta$-transitions as $\gamma$-transitions.)  The $0$-eigenvalue only has
	geometric multiplicity of $g_0 = 1$, so the structure is nondiagonalizable
	for $N>2$. Nevertheless, the generalized eigenvectors are easy to
	construct. The spectral analysis of the cluster structure in (b) suggests
	more general rules that can be gleaned from reading-off eigenvalues from
	digraph clusters; e.g., if a \emph{chain} of $\alpha$'s appears in the
	bisecting path.
	}
\label{fig:DirectedGraphicalClusters}
\end{figure}

\begin{figure*}
\begin{center}
\includegraphics[width=0.75\textwidth]{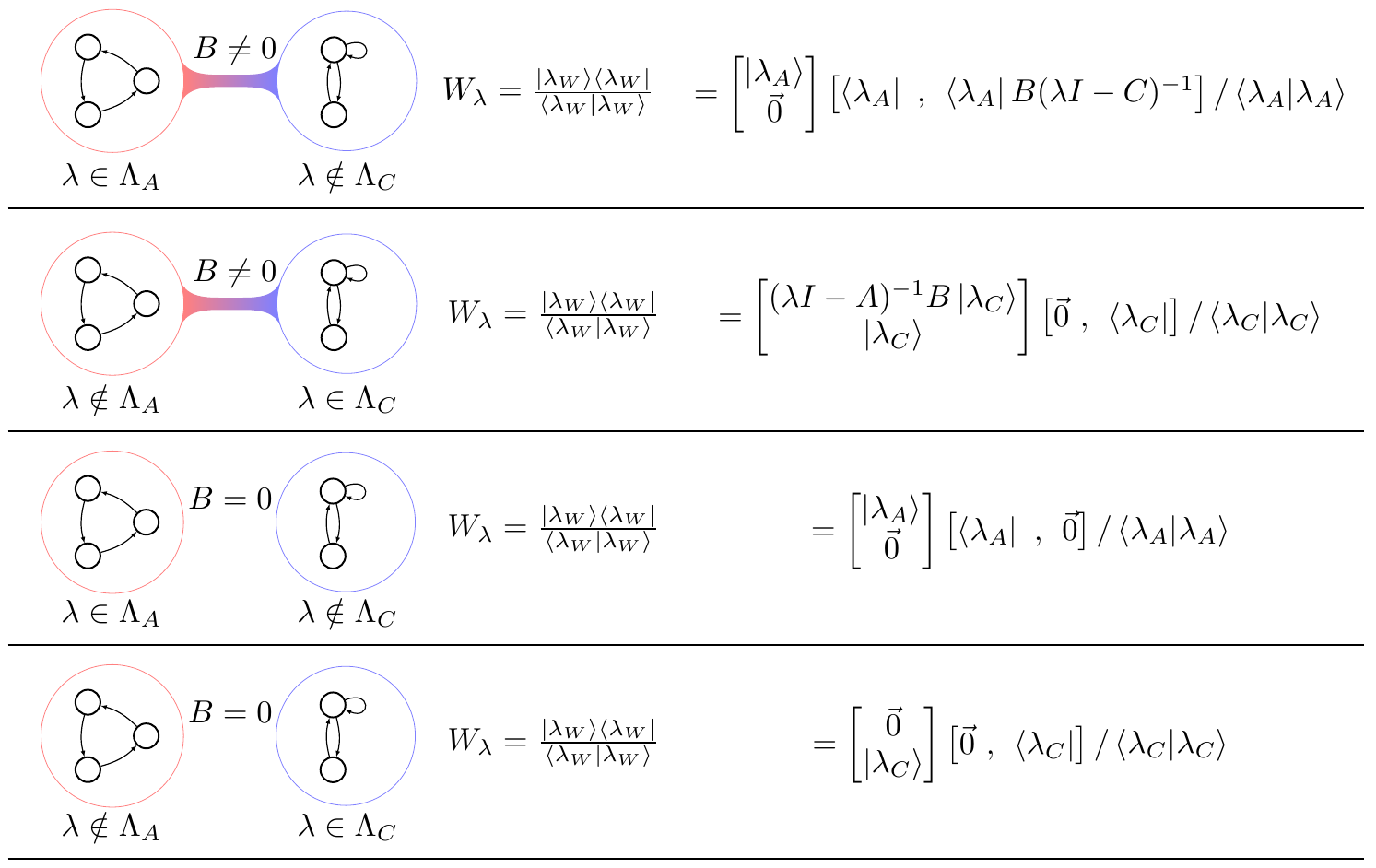}
\end{center}
\caption{Construction of $W$-eigenprojectors $W_\lambda$ from low-level $A$-projectors
	and $C$-projectors, when $W = \begin{bmatrix} A & B \\ \boldsymbol{0} & C
	\end{bmatrix}$. (Recall that $(\lambda I - A)^{-1}$ and $(\lambda I -
	C)^{-1}$ can be constructed from the lower-level projectors.)  For
	simplicity, we assume that the algebraic multiplicity $a_\lambda = 1$ in
	each of these cases.
	}
\label{fig:IterativeProjectorConstruction}
\end{figure*}

\subsection{Row sums} 

If $\matHMM$ is row-stochastic, then by definition: 
\begin{align*}
\matHMM  \ket{\one} =  \ket{\one}
  ~.
\end{align*}
Hence, via the general eigenprojector construction
Eq.~\eqref{eq:ProjectorsViaGenEigenvectors} and the general orthogonality
condition Eq.~\eqref{eq:GenEigenvectorOrthogonality}, we find that:
\begin{align}
\matHMM_\lambda  \ket{\one}= \delta_{\lambda, 1} \ket{\one}
  ~.
\end{align}

This shows that $\matHMM$'s projection operator $\matHMM_1$ is row-stochastic,
whereas each row of every other projection operator must sum to zero.  This can
also be viewed as a consequence of conservation of probability for dynamics
over Markov chains.

\subsection{Expected stationary distribution}

If unity is the only eigenvalue of $\Lambda_T$ on the unit circle, then the
process has no deterministic periodicities. In this case, every initial
condition leads to an stationary asymptotic distribution. The expected
stationary distribution $\pi_\alpha$ from any initial distribution $\alpha$ is: 
\begin{align}
\bra{\pi_\alpha} 
  & = \lim_{L \to \infty}  \bra{\alpha} \matHMM^L  \nonumber \\ 
  & = \bra{\alpha} \matHMM_1 
  ~.
\label{eq:ExpectedStationaryDistr}  
\end{align}
An attractive feature of Eq.~\eqref{eq:ExpectedStationaryDistr} is that it
holds even for nonergodic processes---those with multiple stationary components.

When the stochastic process is ergodic (one stationary component), then $a_1 =
1$ and there is only one stationary distribution $\pi$. The $T_1$ projection
operator becomes:
\begin{align}
T_1 = \ket{\one}\bra{\pi} ~,
\end{align}
even if there are deterministic periodicities. Deterministic periodicities
imply that different initial conditions may still induce different asymptotic
oscillations, according to $\{ T_\lambda : | \lambda | = 1 \}$. In the case of
ergodic processes without deterministic periodicities, every initial condition
relaxes to the same steady-state distribution over the hidden states:
$\bra{\pi_\alpha} = \bra{\alpha} \matHMM_1 \nonumber = \bra{\pi}$ regardless of
$\alpha$, so long as $\alpha$ is a properly normalized probability
distribution.

\begin{table*}
\begin{center}
\begin{tabular}{r l}
\begin{tabular}{r}
Derivatives of cascading $\uparrow$ \\
\phantom{a}\\
Integrals of cascading $\downarrow$ \\
\phantom{Int} \\
\phantom{Int}
\end{tabular}
&
\begin{tabular}{ l || c | c | }
& Discrete time  & \multicolumn{1}{c}{Continuous time}  \\ \hline \hline
Cascading & $\phantom{\Bigl( }\braket{ \cdot | A^L | \cdot} \phantom{\Bigr)}$ &   $\braket{ \cdot | e^{t G} | \cdot}$ \\ \hline
Accumulated transients & $\phantom{\Bigl( } \braket{ \cdot | \left( \sum_L (A - A_1)^L \right) | \cdot} \phantom{\Bigr)}$ &   $\braket{ \cdot | \left( \int (e^{t G} - G_0) \, dt \right) | \cdot}$ \\ 
modulated accumulation & $\phantom{\Bigl( } \braket{ \cdot | \left( \sum_L (z A)^L \right) | \cdot} \phantom{\Bigr)}$ &   $\braket{ \cdot | \left( \int (z e^{G})^{t} \, dt \right) | \cdot}$ \\ \hline
\end{tabular}
\end{tabular}
\end{center}
\caption{Once we identify the hidden linear dynamic behind our questions, most
	are either of the \emph{cascading} or \emph{accumulating} type. Moreover,
	if a complexity measure accumulates transients, the Drazin inverse is
	likely to appear. Interspersed accumulation can be a helpful theoretical
	tool, since all derivatives and integrals of cascading type can be
	calculated, if we know the modified accumulation with $z \in \mathbb{C}$.
	With $z \in \mathbb{C}$, modulated accumulation involves an operator-valued
	$z$-transform. However with $z = e^{i \omega}$ and $\omega \in \mathbb{R}$,
	modulated accumulation involves an operator-valued Fourier-transform.
	}
\label{table:FullQuestionTypes}
\end{table*}
\section{Spectra by inspection}
\label{sec:SpectraByInSpection}

As suggested in Ref.~\cite{Riec16a}, the new results above extend spectral
theory to arbitrary functions of nondiagonalizable operators in a way that
gives a \emph{spectral weighted digraph theory} beyond the purview of spectral
graph theory proper \cite{Cvet98a}. Moreover, this enables new analyses. The
next sections show how spectra and eigenprojectors can be intuited, computed,
and applied in the analysis of complex systems.

\subsection{Eigenvalues}

Consider a directed graph structure with cascading dependencies: one cluster of
nodes feeds back only to itself according to matrix $A$ and feeds forward to
another cluster of nodes according to matrix $B$, which is not necessarily a
square matrix. The second cluster feeds back only to itself according to matrix
$C$. The latter node cluster might also feed forward to another cluster, but
such considerations can be applied iteratively.

The simple situation just described is summarized, with proper index
permutation, by a block matrix of the form: $W = \begin{bmatrix} A & B \\
\boldsymbol{0} & C \end{bmatrix}$. In this case, it is easy to see that:
\begin{align}
\text{det}(W - \lambda I)
  & = \left| 
  \begin{matrix}
  A - \lambda I & B \\ \boldsymbol{0} & C - \lambda I
  \end{matrix}
  \right| \\
  & = | A - \lambda I | \, | C - \lambda I |
  ~.
\end{align}
And so, $\Lambda_W = \Lambda_A \cup \Lambda_C$. This simplification presents an
opportunity to read off eigenvalues from clustered graph structures that often
appear in practice, especially for transient graph structures associated with
transient causal states in \eMs.

Cyclic cluster structures (say, of length $N$ and edge-weights $\alpha_1$
through $\alpha_N$) yield especially simple spectra: 
\begin{align}
\Lambda_A = \Bigl\{
  \bigl( \prod_{i=1}^N \alpha_i \bigr)^{1/N}
  e^{i n 2 \pi / N}
  \Bigr\}_{n=0}^{N-1}
  ~.
\label{eq:cyclic_eig_rule}
\end{align}
That is, the eigenvalues are simply the $N^\text{th}$ roots of the product of
all of the edge-weights. See Fig. \ref{subfig:CyclicCluster}.

Similar rules for reading off spectra from other cluster structures exist.
Although we cannot list them exhaustively here, we give another simple but
useful rule in Fig.~\ref{subfig:DoublyCyclicCluster}. It also indicates the
ubiquity of nondiagonalizability in weighted digraph structures. This second
rule is suggestive of further generalizations where spectra can be read off
from common digraph motifs.

\subsection{Eigenprojectors from graph structure}

We just outlined how clustered directed graph structures yield simplified joint
spectra. Is there a corresponding simplification of the projection operators?
In fact, there is and it leads to an iterative construction of ``higher-level''
projectors from ``lower-level'' clustered components. In contrast to the joint
spectrum though, that completely ignores the feedforward matrix $B$, the
emergent projectors do require $B$ to pull the associated eigencontributions
into the generalized setting. Figure~\ref{fig:IterativeProjectorConstruction}
summarizes the results for the simple case of nondegenerate eigenvalues. The
general case is constructed similarly.

\begin{table*}
\begin{center}
\includegraphics[width=\textwidth]{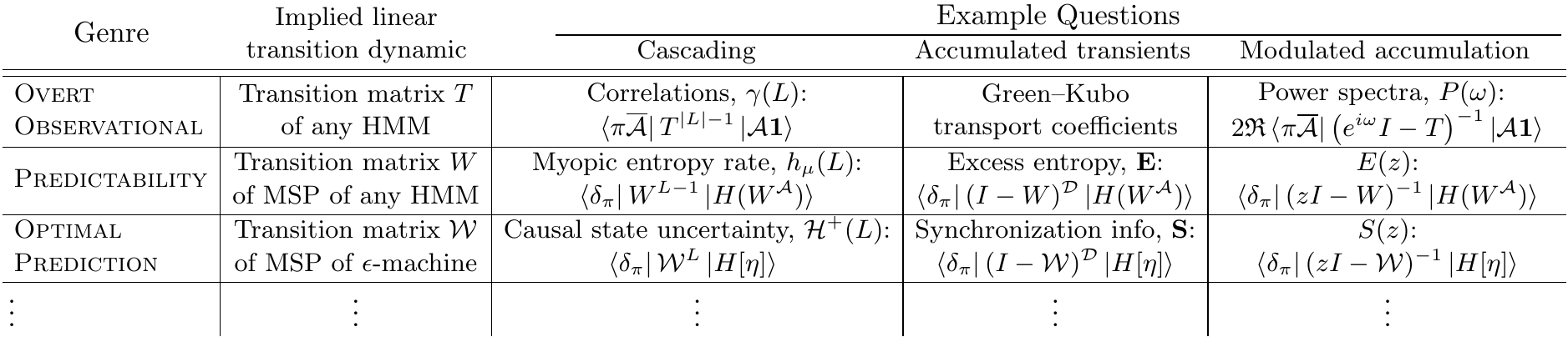}
\end{center}
\caption{Genres of complexity questions given in order of increasing
	sophistication; summary of Part I and a preview of Part II. Each implies a
	different linear transition dynamic.  Closed-form formulae are given for
	several complexity measures, showing the similarity among them down the
	same column. Formulae in the same row have matching bra-ket pairs. The
	similarity within the column corresponds to similarity in the
	time-evolution implied by the question type. The similarity within the row
	corresponds to similarity in question genre.
	}
\label{table:FullClosedFormSolutions}
\end{table*}

The preceding results imply a number of algorithms, both for analytic and
numerical calculations. Most directly, this points to the fact that
eigenanalysis can be partitioned into a series of simpler problems that are
later combined to a final solution. However, in addition to more efficient
serial computation, there are opportunities for numerical parallelization of
the algorithms to compute the eigenprojectors, whether they are computed
directly, say from Eq.~\eqref{eq: proj operators for index-one eigs}, or from
right and left eigenvectors and generalized eigenvectors. Such automation is
useful for applying our analysis to real systems with immense data produced
from very high-dimensional state spaces.

\section{Conclusion}
\label{sec:Conclusion}

Surprisingly, many questions we ask about a structured stochastic nonlinear
process imply a linear dynamic over a preferred hidden state space. These
questions often concern predictability and prediction. To make predictions
about the real world, though, it is not sufficient to have a model of the
world. Additionally, the predictor must synchronize their model to the
real-world data that has been observed up to the present time. This metadynamic
of synchronization---the transition structure among belief states---is
intrinsically linear, but is typically nondiagonalizable.

Recall the organizational tables from the Introduction. After all of the intervening detail, let's consider a more nuanced formulation. We saw that once we frame our questions in terms of the hidden linear transition dynamic, 
complexity measures are usually either of the cascading or accumulation type.  Scalar complexity measures often accumulate only the interesting transient structure that rides on top of the asymptotics. Skimming off the asymptotics led to a Drazin inverse. Modified accumulation turns complexity scalars into complexity functions. This is summarized in Table~\ref{table:FullQuestionTypes} 
and Table~\ref{table:FullClosedFormSolutions}. It is notable that
Table~\ref{table:FullClosedFormSolutions} gives closed-form formulae for many
complexity measures that previously were only expressed as infinite sums over
functions of probabilities.

Let us remind ourselves: Why, in this analysis, were nondiagonalizable dynamics
noteworthy? They are noteworthy since the \emph{metadynamics} of diagonalizable
dynamics are generically nondiagonalizable---typically due to the
zero-eigenvalue subspace that is responsible for the initial, ephemeral epoch
of symmetry collapse. We saw this explicitly with the metadynamics of
transitioning between belief states. However, other metadynamics beyond that
focused on prediction are also generically nondiagonalizable. For example, in
the analysis of quantum compression, crypticity, and other aspects of hidden
structure, the relevant linear dynamic is not the MSP, but is nevertheless a
nondiagonalizable structure that is fruitfully analyzed with the recently
generalized spectral theory of nonnormal operators~\cite{Riec16a}.

Using the appropriate dynamic for common complexity questions and the
meromorphic functional calculus to overcome nondiagonalizability, the sequel
(Part II) goes on to develop closed-form expressions for complexity measures as
simple functions of the corresponding transition dynamic of the implied HMM.

\section*{Acknowledgments}

JPC thanks the Santa Fe Institute for its hospitality. The authors thank Chris
Ellison, Ryan James, John Mahoney, Alec Boyd, and Dowman Varn for helpful discussions. This material is
based upon work supported by, or in part by, the U. S. Army Research Laboratory
and the U. S. Army Research Office under contract numbers W911NF-12-1-0234, W911NF-13-1-0340, and
W911NF-13-1-0390.

\bibliography{chaos}

\end{document}